\pdfoutput=1

\documentclass[onefignum,onetabnum]{siamart171218}

\usepackage{hyperref}

\usepackage{tikz}
\usepackage{pgfplots}
\usepackage[makeroom]{cancel}

\usepackage{lipsum}
\usepackage{amsfonts}
\usepackage{graphicx}
\usepackage{epstopdf}
\usepackage{algorithmic}
\ifpdf
  \DeclareGraphicsExtensions{.eps,.pdf,.png,.jpg}
\else
  \DeclareGraphicsExtensions{.eps}
\fi

\newsiamremark{remark}{Remark}
\newsiamremark{hypothesis}{Hypothesis}
\crefname{hypothesis}{Hypothesis}{Hypotheses}
\newsiamthm{claim}{Claim}

\headers{Multiple Waves Propagate in Random Particulate Materials}{A. L. Gower et. al.}

\title{Multiple Waves Propagate in Random Particulate Materials\thanks{Submitted to the editors October 2018.
\funding{This work was funded EPSRC (EP/M026205/1,EP/L018039/1) and support from the Isaac Newton Institute (EP/K032208/1).
}}}

\author{
  Artur L. Gower\footnotemark[3] \thanks{
    Department of Mechanical Engineering, The University of Sheffield, UK (\email{arturgower@gmail.com}, \url{http://arturgower.github.io}).
  }
  \and
  William J. Parnell\thanks{
    School of Mathematics, University of Manchester, Oxford Road, Manchester M13 9PL, UK.
  }
  \and
   I. David Abrahams\thanks{
    Isaac Newton Institute for Mathematical Sciences, 20 Clarkson Road, Cambridge CB3 0EH, UK
  }
}

\usepackage{amsopn}

\usepackage[colorinlistoftodos,bordercolor=orange,backgroundcolor=orange!20,linecolor=orange,textsize=scriptsize]{todonotes}
\usepackage{soul}

\newcommand{\wrong}[1]{\textcolor{black}{#1}}
\newcommand{\edit}[1]{\textcolor{black}{#1}}
\newcommand{\wieneredit}[1]{\textcolor{black}{#1}}

\newcommand \scatZ {Z}

\newcommand \Ab {\mathcal A}
\newcommand \A [1] {\Ab_#1}
\newcommand \p {p}

\newcommand \reg {\mathcal R}

\newcommand {\nfrac}[1] {\mathfrak n_{#1}}

\newcommand \inc{\mathrm{inc}}

\newcommand{\ensem}[1]{\langle #1 \rangle}

\def\bal#1\eal{\begin{align}#1\end{align}} 
\def\bals#1\eals{\begin{align*}#1\end{align*}}

\renewcommand{\vec}[1]{\boldsymbol{#1}}

\newcommand{\ii}{\textrm{i}}
\newcommand{\ee}{\textrm{e}}

\ifpdf
\hypersetup{
  pdftitle={Multiple Waves Propagate in Random Particulate Materials},
  pdfauthor={A. L. Gower}
}
\fi

\graphicspath{{images/}}

\begin{document}

\maketitle

\begin{abstract}
  \edit{For over 70 years it has been assumed that scalar wave propagation in (ensemble-averaged) random particulate materials can be characterised by a single effective wavenumber. Here, however, we show that there exist many effective wavenumbers, each contributing to the effective transmitted wave field.
   Most of these contributions rapidly attenuate away from boundaries, but they make a significant contribution to the reflected and total transmitted field beyond the low-frequency regime.
   In some cases at least \textit{two} effective wavenumbers have the same order of attenuation. In these cases a single effective wavenumber does not accurately describe wave propagation even far away from boundaries. We develop an efficient method to calculate all of the contributions to the wave field for the scalar wave equation in two spatial dimensions, and then compare results with numerical finite-difference calculations. This new method is, to the authors' knowledge, the first of its kind to give such accurate predictions across a broad frequency range and for general particle volume fractions. }
\end{abstract}




 \section{Introduction}
\edit{Materials comprising small particles, inclusions or defects, randomly distributed inside an otherwise uniform host medium are ubiquitous. Commonly occurring examples include composites, emulsions, dense suspensions, complex gases, polymers and foods. Understanding how electromagnetic, elastic, or acoustic waves propagate through such media is crucial to characterise these materials and also to design new materials that can control wave propagation. For example, we may wish to use wave techniques to \textit{determine} statistical information about the material, e.g.\ volume fraction of particles, particle radius distribution, etc.}

\edit{The exact positions of all particles is usually unknown. The common approach to deal with this, which we adopt here, is to ensemble average over such unknowns. In certain scenarios, such as light scattering~\cite{mishchenko_multiple_2006}, it is easier to measure the average intensity of the wave, but these methods often need the ensemble-averaged field as a first step~\cite{foldy_multiple_1945,tsang_radiative_1987,tsang_dense_2000}.}

\subsection{Historical perspective}
\label{sec:historical}

\edit{The seminal work in this field is Foldy's 1945 paper~\cite{foldy_multiple_1945}, which introduced the \textit{Foldy closure approximation} in order to deduce \textit{a single `effective wavenumber' $k_*$} in the form $k_*= k_0 - \phi g$ where $\phi$ is the \textit{volume fraction} of particles and $g$ is the scattering coefficient associated with a single particle. Foldy introduced the notion of ensemble averaging the field, but the expression deduced for $k_*$ was restricted to dilute dispersions and isotropic scattering. Lax improved on this by incorporating a higher-order closure approximation~\cite{lax_multiple_1951,lax_multiple_1952}, now known as the `\textit{Quasi-Crystalline Approximation}' (QCA), and by including pair-correlation functions, which represent particle distributions. Both QCA and pair-correlations have now been extensively used in multiple scattering theory.
 The most commonly used pair-correlation is '\textit{hole correction}'~\cite{fikioris_multiple_1964}. Both QCA and hole-correction are examples of statistical closure approximations~\cite{adomian_closure_1971,adomian_closure_1979}, which are techniques widely used in statistical physics. For multiple scattering, the accuracy of these approximations has been supported by theoretical~\cite{martin_multiple_2008,martin_estimating_2010}, numerical~\cite{chekroun_time-domain_2012} and experimental~\cite{varadan_multiple_1983,west_comparison_1994} evidence. These approximations also make no explicit assumptions on the frequency range, material properties, or particle volume fraction. We note however that, to our are knowledge, there are no rigorous bounds for the error introduced by these approximations. For a brief discussion on these approximations see~\cite{gower_reflection_2018}. }

\edit{For an overview of the literature on multiple scattering in particulate materials, making use of closure approximations, see the books~\cite{tsang_scattering_2004,martin_multiple_2006,mishchenko_multiple_2006}. We now briefly summarise how calculating effective wavenumbers has evolved since the early work of Foldy and Lax.}

\edit{Over the last 60 years, corrections to the dilute limit have been sought, mainly by expanding in the volume fraction $\phi$ and then attempting to determine the $\mathrm O(\phi^2)$ contribution to $k_*$. Twersky~\cite{twersky_scattering_1962} obtained an expression for this contribution as a function of $f(\pi-2\theta_\inc)$ and $f(0)$, where $\theta_\inc$ is the angle of incidence of an exciting plane wave, see \Cref{fig:coordinates}, and $f(\theta)$ is the far field scattering pattern from one particle~\cite{linton_multiple_2005}.
 The dependence on $\theta_\inc$ implies that $k_*$ depends on the angle of incidence, which is counter-intuitive. Waterman \& Truell~\cite{waterman_multiple_1961} obtained the same expression as Twersky but with $\theta_\inc=0$. However, \cite{waterman_multiple_1961} used a `slab pair-correlation function' that (theoretically) limits the validity of their approach to dilute dispersions (small $\phi$), see~\cite{layman_interaction_2006} for comparisons with experiments, and see~\cite{aristegui_effective_2007} for a discussion in two dimensions. Extensions that incorporate the hole-correction pair-correlation function were described by Fikioris \& Waterman~\cite{fikioris_multiple_1964}. The Waterman \& Truell expressions for three-dimensional (3D) elasticity are written down in~\cite{yang_dynamic_2003,parnell_effective_2010}. Work in 2D elasticity using QCA was reported by~\cite{bose_longitudinal_1973}.
 Lloyd and Berry~\cite{lloyd_wave_1967} calculated the $\mathrm O(\phi^2)$ contribution by including both QCA and hole correction for the scalar wave equation, although the language used stemmed from nuclear physics. More recently, \cite{linton_multiple_2005,linton_multiple_2006} re-derived the Lloyd \& Berry formula for the effective wavenumber without appealing to the so-called extinction theorem used in many previous papers, such as~\cite{varadan_multiple_1985}, and without recourse to `resumming series'. The work was then extended in order to calculate effective reflection and transmission in~\cite{martin_multiple_2011}. Gower et al.~\cite{gower_reflection_2018} subsequently extended this result to model multi-species materials, i.e.\ to account for polydisperse distributions.
}

\edit{Other related work on effective wavenumbers and attenuation include: comparing the properties of single realisations to that of effective waves~\cite{rupprecht_calculation_2019,bennetts_spectral_2013,bennetts_absence_2015,montiel_evolution_2015}, and effective waves in polycrystals~\cite{sha_correlation_2018,weaver_diffusivity_1990} such as steel and ceramics. The polycrystal papers use a similar framework to waves in particulate materials, except they assume weak scattering which excludes multiple scattering.
}

\subsection{Overview of this paper}

\edit{A common \textit{assumption} used across the field of random particulate materials, including those mentioned above, is to assume there exists a \textit{single, unique, complex effective wavenumber} $k_*$ that characterises the material. For example, for an incident wave $\ee^{\ii k x- \ii\omega t}$, of fixed frequency $\omega$, exciting a half-space (see \Cref{fig:coordinates}) filled with particles, the tacit assumption is that the ensemble averaged wave $\ensem{u(x)}$ travelling inside the particulate material takes the form
\begin{align}
\ensem{u(x)} = a \ee^{\ii k x-\ii\omega t} + b_* \ee^{\ii k_*x- \ii\omega t}.
\end{align}
See~\cite{martin_multiple_2011} for a brief derivation. This assumption has been widely used in acoustics~\cite{linton_multiple_2005,linton_multiple_2006,martin_multiple_2006,dubois_coherent_2011}, elasticity ~\cite{varadan_multiple_1978,norris_scattering_1986,norris_effective_2012,pinfield_thermo-elastic_2014,conoir_effective_2010}  (including thermo-viscous effects), electromagnetism~\cite{varadan_coherent_1979,varadan_multiple_1983,tishkovets_scattering_2011}, and even quantum~\cite{sheng_introduction_2006} waves. For example, it is a key step in deducing radiative transfer equations from first principles~\cite{mishchenko_vector_2002,mishchenko_first-principles_2016}.}

\edit{In this work, we show however that there \emph{does not exist a single, unique} effective wavenumber. Instead an infinite number of effective wavenumbers $k_1,k_2,k_3,....$ exist, so that the average field inside the particulate material takes the form
\begin{align}
\langle u(x) \rangle = a e^{ikx-i\omega t} + \sum_{p=1}^{\infty}b_p e^{ik_p x-i\omega t}.
\label{eqn:transmitted_effective}
\end{align}
In many scenarios, the majority of these waves are highly attenuating, i.e.\ $k_p$ has a large imaginary part for $p>1$. In these cases, the least attenuating wavenumber $k_1$ will dominate the transmitted field in $\langle u(x)\rangle$, and $k_1$ will often be given by classical multiple scattering theory, as discussed in \Cref{sec:historical}. However, these other effective waves can still have a significant contribution to the reflected (backscattered) wave from a random particulate material, especially at higher frequencies and beyond the low volume fraction $\phi$ limit. Furthermore, there are scenarios where there are at least two effective wavenumbers, say $k_1$ and $k_2$, with the same order of attenuation. In these cases using only one effective wavenumber, $k_1$ or $k_2$, is insufficient to accurately calculate $\langle u(x)\rangle$, even for $x$ far away from the interface between the homogeneous and particular materials.}

\edit{We examine the simplest case that exhibits these multiple waves: two spatial dimensions $(x,y)$ for the scalar wave equation, and consider particles placed in the half-space $x>0$, which reflects incoming waves. We not only demonstrate that there are multiple effective wavenumbers, but we also use them to develop a highly accurate method to calculate $\ensem{u(x)}$ and the reflection coefficient. This method agrees extremely well with numerical solutions, calculated using a finite difference method, but is more efficient. We provide software~\cite{gower_effective_waves.jl:_2018} that implements the methods presented and reproduce the results of this paper.}

\edit{In a separate paper~\cite{gower_proof_2019}, we develop a \textit{proof} that~\eqref{eqn:transmitted_effective} is the analytical solution for the ensemble averaged wave. However, the proof in~\cite{gower_proof_2019} is not constructive, in contrast to the work presented here, where we present a method that determines all effective waves~\eqref{eqn:transmitted_effective}.}

We begin by deducing the governing equation for ensemble averaged waves in \Cref{sec:governing}. In \Cref{sec:effective-waves} we then show that multiple effective wavenumbers exist. To calculate these effective wavenumbers, we need to match them to the field near the interface at $x=0$, which leads us to develop a discrete solution in \Cref{sec:finite_difference}. The discrete solution also serves as the basis for a numerical method, which we use later as a benchmark. In \Cref{sec:matching} we develop the Matching method, which incorporates all of the effective waves. \edit{In \Cref{sec:methods} we summarise the fields and reflection coefficients calculated by the Matching method, the numerical method, and extant methods used in the literature. We subsequently compare their results in \Cref{sec:experiments}. In~\Cref{sec:conclusions} we summarise the main results of the paper and discuss future work.}

\section{Ensemble averaged multiple scattering}
\label{sec:governing}

Consider a region $\reg$ filled with $N$ particles or inclusions that are uniformly distributed. The field $u$ is governed by the scalar wave equations:
\begin{align}
  &\nabla^2 u + k^2 u = 0, \quad \text{(in the background material)}, \\
  &\nabla^2 u + k^2_o u = 0, \quad \text{(inside a particle)},
\end{align}
where $k$ and $k_o$ are the real wavenumbers of the background and inclusion materials, respectively. We assume all particles are the same, except for their position and rotation about their centre, for simplicity. For a distribution of particles, or multi-species, see~\cite{gower_reflection_2018}.

\begin{figure}[htbp]
  \centering
  \includegraphics[width=0.65\linewidth]{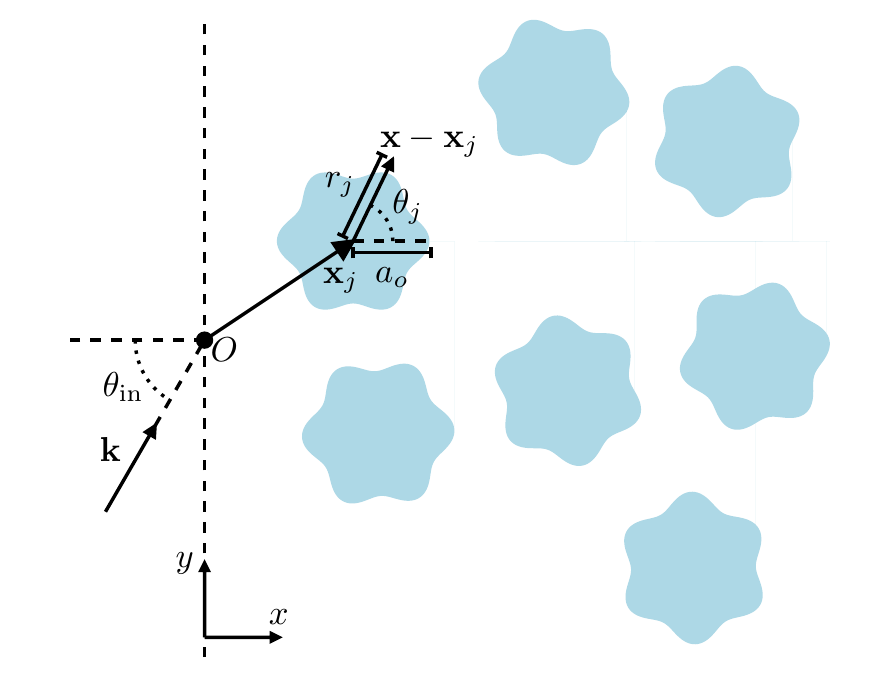}
  \caption{Coordinates for particles with the origin $O$. The particles are only placed in $x>0$, that is, to the right of the dashed line. The vector $\mathbf k = k (\cos \theta_\inc, \sin \theta_\inc)$ shows the direction of the incident plane wave. }
  \label{fig:coordinates}
\end{figure}

In two dimensions, any incident wave\footnote{Equation~\eqref{eqn:incident_j} assumes that the incident wave originates outside of the of the $j$-th particle, which is normally the case.} $v_j$ and scattered wave $u_j$ can be written in the form
\begin{align}
  & v_j(r_j,\theta_j) = \sum_{n=-\infty}^\infty v_{nj} \mathrm J_{n}(k r_j) \ee^{\ii n \theta_j},
  \label{eqn:incident_j}
  \\
  & u_j(r_j,\theta_j) = \sum_{n=-\infty}^\infty u_{nj} \mathrm H_{n}(k r_j) \ee^{\ii n \theta_j},
\label{eqn:scattered_j}
\end{align}
with $(r_j, \theta_j)$ being the polar coordinates of $\mathbf x - \mathbf x_j$, and $\mathbf x_j = (x_j,y_j)$ a vector pointing to the centre of the $j$-th particle, from some suitable origin, and $\mathbf x$ is any vector, see \Cref{fig:coordinates}. The $\mathrm J_n$ and $\mathrm H_n$ are respectively Bessel and Hankel functions of the first kind. The representation \eqref{eqn:scattered_j} is valid when $r_j$ is large enough for $(r_j,\theta_j)$ to be outside of the $j$-th particle for all $\theta_j$. For example, in \Cref{fig:coordinates} this distance would be $r_j > a_o$.

The T-matrix is a linear operator, in the form of an infinite matrix, such that
\begin{equation}
  u_{\wrong{nj}} = \sum_{m=-\infty}^\infty T_{nm}(\edit \tau_j) v_{mj} \quad \edit{ \text{for} \;\; n= -\infty, \ldots, \infty \;\; \text{and} \;\; j =1,\ldots, N},
  \label{eqn:T-matrix}
\end{equation}
where we recall that $N$ is the number of particles. The angle $\edit \tau_j$ gives the particles rotation about their centre $\vec x_j$. \edit{Allowing particles to have different rotations, and assuming all $\tau_j \in [0, 2 \pi]$ to be equally likely, will lead to ensemble average equations that are equivalent to the equations for circular particles~\cite{mishchenko_t-matrix_1996}.} This matrix $\mathbf T$ exists when scattering is a linear operation (elastic scattering), and can accommodate particles with a large variety of shapes and properties~\cite{ganesh_far-field_2010,ganesh_algorithm_2017,waterman_symmetry_1971}; it is especially useful for multiple scattering~\cite{varadan_multiple_1978,mishchenko_t-matrix_1996,jondea_multiplescattering.jl:_2017}.

The T-matrix also accounts for the particle's boundary conditions. For instance, if $u$ represents pressure, $\rho$ and $c$ are the background density and wave speed, the particles are circular with density $\rho_o$, sound-speed $c_o$, and radius $a_o$, then continuity of pressure and displacement across the particle's boundary~\cite[Section IV A]{linton_multiple_2005}, yields
\begin{equation}
  T_{nm} = - \delta_{nm} \scatZ^m_o, \;\; \text{with} \;\; Z^m_o = \frac{q_o J_m' (k a_o) J_m (k_o a_o) - J_m (k a_o) J_m' (k_o a_o) }{q_o H_m '(k a_o) J_m(k_o a_o) - H_m(k a_o) J_m '(k_o a_o)},
  \label{eqn:circular_t-matrix}
\end{equation}
where $q_o = (\rho_o c_o)/(\rho c)$ and $k_o = \omega/c_o$. In this case the T-matrix is independent of the rotation $\edit \tau_j$.

In this paper we consider the incident plane wave
\begin{equation}
  u_\inc(x,y) = \ee^{\ii k (x \cos \theta_\inc + y \sin \theta_\inc)} \quad \text{for} \;\; \theta_\inc \in (-\pi/2,\pi/2),
  \label{eqn:incident}
\end{equation}
which excites $N$ particles, resulting in scattered waves of the form~\eqref{eqn:scattered_j}. See \Cref{fig:coordinates} for an illustration. The total wave $u$, measured outside of all particles at $\mathbf x = (x,y)$, is the sum of all scattered waves plus the incident wave:
\begin{equation}
  u(x,y) = u_\inc(x,y) + \sum_{j=1}^N u_j(r_j,\theta_j).
  \label{eqn:total_u}
\end{equation}

\edit{
To reach an equation for the coefficients $u_{nj}$ we write the total wave field incident on the $j$-th particle $v_j$~\eqref{eqn:incident_j} as a combination of the incident wave plus the waves scattered by the other particles: $v_j(r_j,\theta_j) = u_\inc(x,y) + \sum_{i\not = j} u_i(r_i,\theta_i)$. By then applying the Jacobi-Anger expansion to $u_\inc(x,y)$, using Graf's addition theorem~\cite{martin_multiple_2006,abramowitz_handbook_1966}, multiplying both sides by  $T_{qn}$, summing over $n$, and then using~\eqref{eqn:T-matrix}, we obtain
\begin{multline}
  u_{qj} = u_\inc (x_j,y_j)\sum_{n=-\infty}^\infty T_{qn}(\edit \tau_j) \ee^{\ii n (\pi/2 - \theta_\inc)}
  \\
  + \sum_{i\not =j}\sum_{m,n=-\infty}^\infty u_{mi} T_{qn}(\edit \tau_j) F_{m-n}(k\vec x_i - k\vec x_j),
  \label{eqn:FiniteMultipleScattering}
\end{multline}
for all integers $q$ and $j=1,2, \ldots, N$, where
}
\begin{equation}
   F_{n}(\mathbf X) = (-1)^{n}\ee^{\ii n \Theta} H_{n}(R),
  \label{eqn:integral_kernel}
\end{equation}
and $(R,\Theta)$ are the polar coordinates of $\mathbf X$.

\subsection{Ensemble averaging}
In practice the exact position of the particles is unknown, so rather than determine the scattering from an exact configuration of particles, we ensemble average the field $u$ over all possible particle rotations and positions in $\reg$. Sensing devices also naturally perform ensemble averaging due to their size or from time averaging~\cite{mishchenko_multiple_2008}.  See~\cite{foldy_multiple_1945,parnell_effective_2010,gower_reflection_2018} for an introduction to ensemble-averaging of multiple scattering.

For simplicity, we assume the particle positions are independent of particle rotations, so that the probability of the particles being centred at $\mathbf x_1,\mathbf x_2, \ldots, \mathbf x_N$, is given by the probability density function $\p(\mathbf x_1, \mathbf x_2,\ldots, \mathbf x_N)$. Hence, it follows that
\begin{equation}
  \int \p(\mathbf x_1) \mathrm d \mathbf x_1 = \int \int \p(\mathbf x_1, \mathbf x_2) \mathrm d \mathbf x_1 \mathrm d \mathbf x_2 = \ldots = 1,
\end{equation}
where each integral is taken over $\reg$.
Further, we have
\begin{equation}
  \p(\mathbf x_1, \ldots, \mathbf x_N) = \p(\mathbf x_j) \p(\mathbf x_{1}, \ldots, \mathbf x_N|\mathbf x_j),
  \label{eqns:conditional_probj}
\end{equation}
where $\p(\mathbf x_1, \ldots, \mathbf x_N|\mathbf x_j)$ is the conditional probability density of having particles centred at $\mathbf x_1, \ldots, \cancel{\mathbf x_j}, \ldots, \mathbf x_N$ (not including $\mathbf x_j$), given that the $j$-th particle is centred at $\mathbf x_j$. Given some function $F(\mathbf x_1, \ldots, \mathbf x_N)$, we denote its {\it ensemble average} (over particle positions) by
\begin{equation}
  \ensem F  = \int\ldots \int F(\mathbf x_1, \ldots, \mathbf x_N) \p(\mathbf x_1, \ldots, \mathbf x_N) \mathrm d\mathbf x_{1} \ldots \mathrm d\mathbf x_{N}.
\end{equation}
If we fix the location of the $j$-th particle, $\mathbf x_j$, and average over the positions of the other particles, we obtain a {\it conditional average} of $F$ given by
\begin{equation}
  \ensem{F}_{\mathbf x_j} = \int\ldots \int F(\mathbf x_1, \ldots, \mathbf x_N) \p( \mathbf x_1, \ldots, \mathbf x_N|\mathbf x_j)  \mathrm d\mathbf x_{1} \ldots \cancel{\mathrm d \mathbf x_{j}} \ldots \mathrm d\mathbf x_{N},
\end{equation}

 We assume that one particle is equally likely to be centred anywhere in $\reg$, when the position of the other particles is unknown:
\begin{equation}
  \p(\mathbf x_j) = \frac{\nfrac {}}{N}, \quad \text{for} \;\; \mathbf x_j \in \reg,
  \label{eqn:p_xj}
\end{equation}
 where we define the number density $\nfrac {} = N/|\reg|$ and the area of $\reg$ as $|\reg|$.

Using the above we can express $\ensem{u(x,y)}$, for $(x,y)$ outside of the region $\reg$, by taking the ensemble average of both sides of~\eqref{eqn:total_u} to obtain
\bal
  \ensem{u(x,y)} &= u_\inc(x,y) + \sum_{j=1}^N \int_\reg \ensem{u_j(r_j,\theta_j)}_{\mathbf x_j} \p(\mathbf x_j) \mathrm d \mathbf x_j
  \\
  &= u_\inc(x,y) +  \nfrac {} \int_\reg \ensem{u_1(r_1,\theta_1)}_{\mathbf x_1} \mathrm d\mathbf x_1, \quad \edit{\text{for}\;\; \mathbf x \not \in \reg},
\label{eqn:AverageWave}
\eal
where we assumed that all particles are identical (apart from their position and rotation). We also used equations~(\ref{eqns:conditional_probj},\ref{eqn:p_xj}) \edit{and averaged both sides over particle rotations}. Using the scattered field~\eqref{eqn:scattered_j}, we then reach
\begin{equation}
  \ensem{u_1(r_1,\theta_1)}_{\mathbf x_1} = \sum_{n=-\infty}^\infty \ensem{u_{n1}}_{\mathbf x_1} H_n(k r_1) \ee^{\ii n \theta_1}.
  \label{eqn:AverageWaveCond}
\end{equation}
The simplest scenario is the limit when the particles occupy the half-space $x_1 > 0$~\cite{linton_multiple_2005}\edit{, that is $\reg = \{(x,y) | x > 0\}$ }. We focus on this case, although the method we present can be adapted to any region $\reg$. In the limit of $\reg$ tending to a half-space, we let $N \to \infty$ while $\nfrac {}$ remains fixed. Due to the symmetry between the incident wave~\eqref{eqn:incident} and the half-space $x_1 > 0$, the \edit{field $\ensem{u_{n1}}_{\mathbf x_1}$ has a translational symmetry along $y_1$}, which allows us to write~\cite{gower_reflection_2018}
\begin{equation}
  \ensem{u_{n1}}_{\mathbf x_1} = \Ab_n(kx_1) \ee^{\ii k y_1 \sin \theta_\inc}.
  \label{eqn:u_to_A}
\end{equation}

For step-by-step details on deriving a governing equation for $\Ab_n(kx_1)$, see~\cite{linton_multiple_2005,conoir_effective_2010,gower_reflection_2018}. Here we only give an overview. First multiple $p(\vec x_2,\ldots,\vec x_N|\vec x_1)$ on both sides of equation~\eqref{eqn:FiniteMultipleScattering}, set $j=1$, ensemble average over all particle rotations\footnote{Assuming that every particle is equally and independently likely to be rotated by any angle $\edit \tau_j$, which makes the ensemble-averaged T-matrix diagonal~\cite{varadan_scattering_1979,mishchenko_t-matrix_1996}.}
 and particle positions in $x > 0$, then use the statistical assumptions \emph{hole correction}\footnote{The assumption hole correction is not appropriate for long and narrow particles. More generally, the method we present can be applied to any pair correlations that depend only on inter-particle distance.} and the quasicrystalline approximation, to reach the system:
\begin{multline}
   \sum_{n=-\infty}^\infty \nfrac {} \int_{\stackrel{x_2 > 0 }{\|\mathbf x_1 - \mathbf x_2 \| > a_{12}}} T_{m} \Ab_n(kx_2) \ee^{\ii k (y_2 -y_1) \sin \theta_\inc} F_{n-m}(k\mathbf x_2 - k \mathbf x_1)  \mathrm d \mathbf x_2
\\
  \wrong{-} \Ab_m(kx_1) + \ee^{\ii k x_1 \cos \theta_\inc}
  T_{m} \ee^{\ii m ( \pi/2 - \theta_\inc )} = 0, \quad \text{for} \quad x_1 >0,
  \label{eqn:ensemAsystem}
\end{multline}
where $T_{m}\delta_{mq} = (2 \pi)^{-1}\int_{0}^{2 \pi} T_{mq}(\edit \tau) \mathrm d \edit \tau$, $\delta_{mq} = 1$ if $m=q$ and $0$ otherwise,
 and $a_{12}$ is the minimum allowed distance between the centre of any two particles. That is, $a_{12}$ is at least twice the radius for circular particles. \edit{For the case shown} in \cref{fig:coordinates} we could choose $a_{12} = 2 a_o$. This minimum distance $a_{12}$ guarantees that particles do not overlap.

When the T-matrix $\mathbf T$ is known, we can determine the field $\Ab_n$ from the system~\eqref{eqn:ensemAsystem}. The aim of this paper is to efficiently solve for $\Ab_n$ and in the process reveal that $\Ab_n$ is composed of a series of effective waves.

For the rest of the paper we employ the non-dimensional variables
\begin{align}
  X_1 = k x_1, \quad X_2 = k x_2,  \quad R_o \gamma = k a_{12},  \quad \phi = \pi \nfrac {} \frac{R_o^2}{k^2}  = \pi \nfrac {} \frac{a_{12}^2}{\gamma^2},
  \label{eqns:non-dimensional}
\end{align}
where $R_o$ is the particles' non-dimensional maximum radius (in \Cref{fig:coordinates} $R_o = k a_o$), $\gamma \geq 2$ a chosen closeness constant, with $\gamma = 2$ implying that particles can touch, and $\phi$ is the particle volume fraction\footnote{For non-circular particles, $\phi$ is slightly larger than the actual particle volume fraction because we use the outer radius $R_o$ ($a_o$ in \Cref{fig:coordinates}) instead of the appropriate average radius.}. Using non-dimensional parameters helps to formulate robust numerical methods and to explore the parameter space.

\section{Effective waves}
\label{sec:effective-waves}
An elegant way to approximate $\Ab_n$ is to assume it is a plane wave of the form~\cite{martin_multiple_2006}
\begin{equation}
  \Ab_n(X) = \ii^n \ee^{-\ii n \edit \varphi} A_n \ee^{\ii X K \cos \edit \varphi} \quad \text{for} \;\; X > \bar X,
  \label{eqn:effective_wave}
\end{equation}
where $K$ is the non-dimensional effective wavenumber ($k K$ is the dimensional effective wavenumber), with Im $K \geq 0$ to be physically reasonable, the factor $\ii^n \ee^{-\ii n \edit \varphi}$ is for later convenience, and $\bar X$ is a length-scale we will determine later. We also restrict the complex angle $\varphi$ by imposing that $-\pi/2 < \mathrm{Re}\,\varphi < \pi/2$ and using
\begin{equation}
  K \sin \edit \varphi = \sin \theta_\inc,
  \label{eqn:Snells}
\end{equation}
 which is due to the translational symmetry of equation~\eqref{eqn:ensemAsystem} in $y_1$, see~\cite[Equation (4.4)]{gower_reflection_2018}. This relation is often called Snell's law.

As the material has been homogenised, it is tempting to make assumptions that are valid for homogeneous materials, such as assuming that only one plane wave~\eqref{eqn:effective_wave} is transmitted into the material. When the particles are very small in comparison to the wavelength, this is asymptotically correct~\cite{parnell_multiple_2010}, but in all other regimes this is not valid, especially close to the edge $\bar X =0$, as we show below.


By substituting the ansatz~\eqref{eqn:effective_wave} into~\eqref{eqn:ensemAsystem}, using~\eqref{eqns:non-dimensional} (see~\cref{app:effective_waves} for details) and by restricting $X_1 > \bar X + \gamma R_o$, we obtain
\begin{align}
  & \wrong{\sum_{n=-\infty}^\infty} M_{mn}(K) A_n  =0, \quad
  M_{mn}(K) = \wrong{-}R^2_o \delta_{mn} + 2 \phi T_{m} \frac{ \mathcal N_{n-m}(K)}{1 - K^2},
  \label{eqn:K1_dispersion}
\\
  & 2 \phi \sum_{n = -\infty}^\infty  \ee^{\ii n  \theta_\inc}
  A_n  \ee^{-\ii n \edit \varphi}  \frac{\ee^{\ii (K \cos \edit \varphi - \cos \theta_\inc){\bar X}}}{K \cos \edit \varphi - \cos \theta_\inc}
  =   \ii \pi R_o^2 \cos \theta_\inc + g(\bar X),
  \label{eqn:extinction}
\\
  & g(\bar X) =  2 \phi \sum_{n = -\infty}^\infty  \ee^{\ii n  \theta_\inc}   (-\ii)^{n-1} \int_{0}^{{\bar X}} \A n(X_2)  \ee^{-\ii X_2 \cos \theta_\inc } \mathrm d X_2,
\end{align}
where
\begin{equation}
  \mathcal N_{n}(K) = \wrong{\gamma} R_o (H_n'(\gamma R_o)J_n(\gamma K R_o) - K H_n(\gamma R_o) J_n'( \gamma K R_o)),
\end{equation}
and~\eqref{eqn:extinction} is often called the extinction theorem, though we will refer to it as the extinction equation.

\begin{figure}[h!]
  \centering
  \begin{tikzpicture}
  \node[inner sep=0pt] at (0,0)
      {  \includegraphics[width=0.46\linewidth]{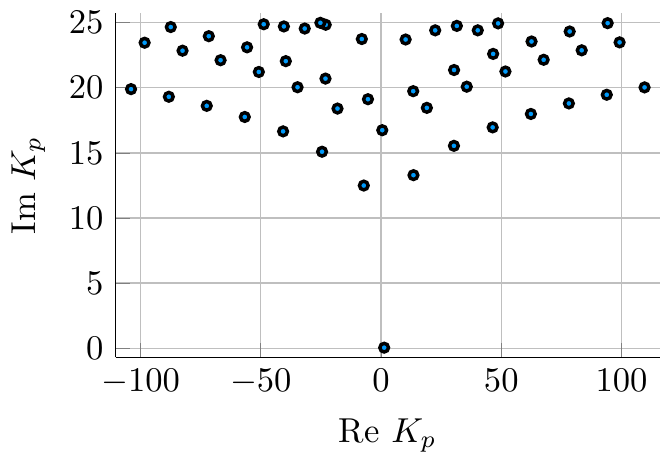}};
  \node[inner sep=0pt] at (6.0,0)
      {  \includegraphics[width=0.46\linewidth]{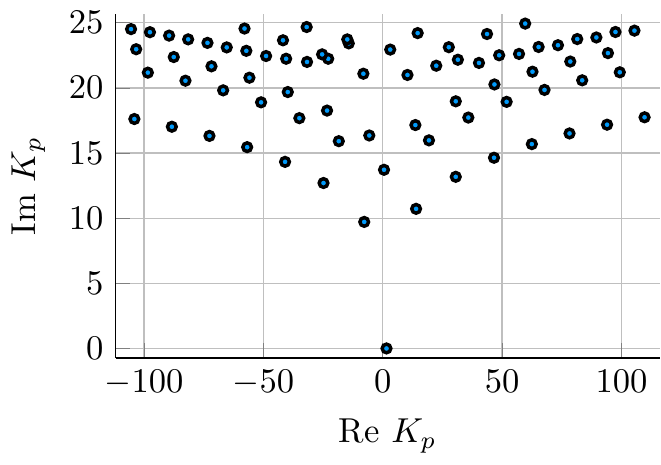}};
  \node[inner sep=0pt] at (0.0,-4.0)
      {  \includegraphics[width=0.46\linewidth]{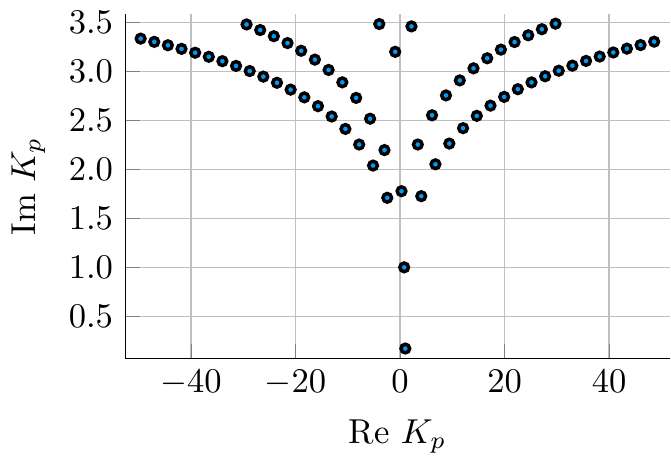}};
  \node[inner sep=0pt] at (6.0,-4.0)
      {  \includegraphics[width=0.46\linewidth]{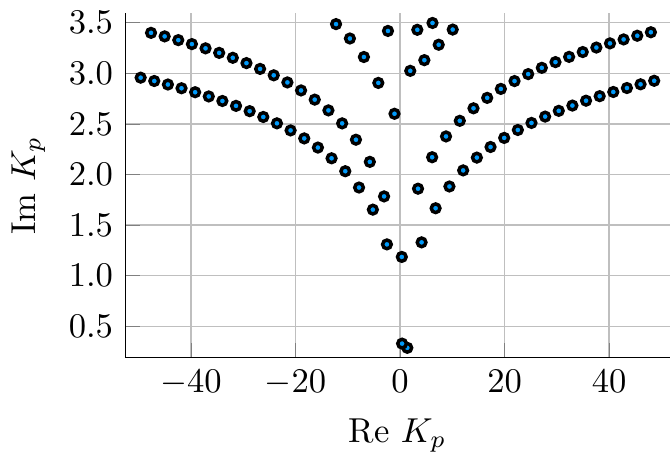}};
  \node at (0.4,2.4) {\Large $\phi = 10\%$};
  \node at (6.4,2.4) {\Large $\phi = 25\%$};
  \node[rotate=90] at (-3.5,0.5) {\Large $R_o = 0.1$};
  \node[rotate=90] at (-3.5,-3.5) {\Large $R_o = 1.2$};
  \end{tikzpicture}
\caption{An example of the effective wavenumbers $K_1,K_2,\ldots$ that satisfy equation~\eqref{eqn:Kp_dispersion}${}_2$. The particles chosen are moderately strong scatterers, with T-matrix~\eqref{eqn:circular_t-matrix}, parameters $k=1$, $k_o = 2.0$, $c_o=\rho_o=0.5$, $c = \rho=1.0$, and the non-dimensional radius $R_o$ is $0.2$ for the top two graphs and $1.2$ for the bottom two graphs. Note that the bottom right graph shows two wavenumbers, almost on top of each other, both with imaginary part less than $0.5$.}
  \label{fig:wavenumbers}
\end{figure}
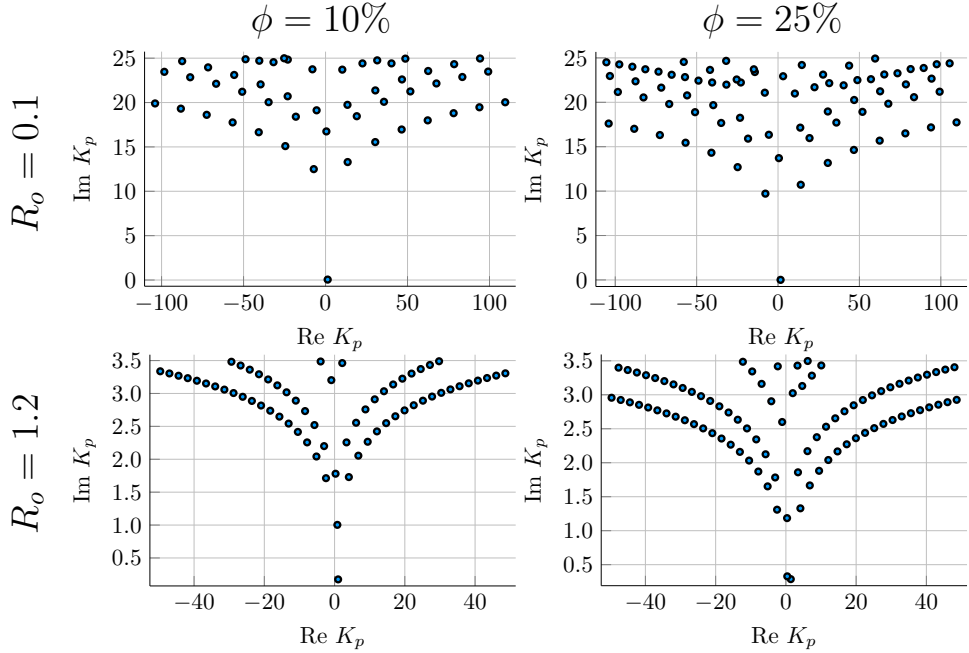

Using~\eqref{eqn:K1_dispersion} we can calculate $K$ by solving
\begin{equation}
\det (M_{mn}(K)) =0,
\label{eqn:detM}
\end{equation}
then \textbf{the standard approach} to calculate $A_n$ is to use \eqref{eqn:K1_dispersion}${}_1$ and \eqref{eqn:extinction} and take $\bar X =0$, which avoids the need to know $\A n$ or to calculate $g(\bar X)$.
It is commonly assumed that there is only one viable $K$, when fixing all the material parameters, including the incident wavenumber $k$. However, in general~\eqref{eqn:detM} admits many solutions, which we denote as $K = K_1$, $K_2$, $\ldots$, see  \Cref{fig:wavenumbers} for some examples. \edit{We order these wavenumbers so that Im $K_p$ increases with $p$.}
There is no reason why these wavenumbers are not physically viable. Therefore we write $\Ab_n$ as a sum of effective waves:
\begin{equation}
\Ab_n(X) = \ii^n \sum_{p=1}^P \ee^{-\ii n \edit \varphi_p} A_n^p \ee^{\ii X K_p \cos \edit \varphi_p} \quad \text{for} \;\; X > \bar X,
\label{eqn:effective_waves}
\end{equation}
\wieneredit{where there are an infinite number of these effective wavenumbers~\cite{gower_proof_2019}, but to reach an approximate method we need only a finite number $P$. Technically, \eqref{eqn:effective_waves} is a solution to~\eqref{eqn:ensemAsystem} for $X>0$, that is, we could take $\bar X =0$. However, in this case, we found that close to $X=0$ a very large number of effective waves $P$ would be required to achieve an accurate solution. This is why we only use the sum of plane waves~\eqref{eqn:effective_waves} for $X > \bar X >0$. }

\edit{One of these effective wavenumbers, in most cases the lowest attenuating $K_1$, can be calculated using an asymptotic expansion for low $\phi$~\cite{linton_multiple_2005}, and assuming it is a perturbation away from $1$ (the background wavenumber).
}

Substituting~\eqref{eqn:effective_waves} into~\eqref{eqn:ensemAsystem} leads to the same dispersion equations~\eqref{eqn:K1_dispersion} and~\eqref{eqn:detM}, but with $K_1$ and $A^1_n$ replaced with $K_p$ and $A^p_n$, which leads to
\begin{equation}
  \wrong{\sum_{n=-\infty}^\infty} M_{mn}(K_p) A^p_n  =0 \quad \text{and}\quad \det (M_{mn}(K_p)) = 0,
   \label{eqn:Kp_dispersion}
\end{equation}
 while for the extinction equation~\eqref{eqn:extinction} we need to substitute $K$ for $K_p$, $\edit \varphi$ for $\edit \varphi_p$, and then sum over $p$ only on the left-hand side to arrive at
\begin{align}
& 2 \phi \sum_{p=1}^P \sum_{n = -\infty}^\infty
A^p_n  \ee^{\ii n ( \theta_\inc -\edit \varphi_p) }  \frac{\ee^{\ii (K_p \cos \edit \varphi_p- \cos \theta_\inc){\bar X}}}{K_p \cos \edit \varphi_p- \cos \theta_\inc}
=   \ii \pi R_o^2 \cos \theta_\inc + g(\bar X),
\label{eqn:extinction_P}
\end{align}
for details see~\Cref{app:effective_waves}. The question now arises: how do we calculate the unknowns $A^p_n$? Once each $K_p$ and $\edit \varphi_p$ are determined from \eqref{eqn:Kp_dispersion}${}_2$ and \eqref{eqn:Snells}, then~\eqref{eqn:Kp_dispersion}${}_1$ can be used to write the vector $ \mathbf A^p = [\ldots, A^p_{-n}, A^p_{1-n}, \ldots, A^p_{n-1}, A^p_{n}, \ldots]$ in the form
\begin{equation}
  \mathbf A^p = \alpha^p \mathbf a^p  \quad \text{and} \quad \vec \alpha = [\alpha^1, \alpha^2, \ldots],
  \label{eqn:A_eigenvector}
\end{equation}
 where the $\mathbf a^p$ are determined from~\eqref{eqn:Kp_dispersion}${}_1$.
   However, only equation~\eqref{eqn:extinction_P} remains to determine the vector $\vec \alpha$. As there is more than one effective wave, $P>1$, equation~\eqref{eqn:extinction_P} is not sufficient to determine $\vec \alpha$. This is because satisfying~\eqref{eqn:Kp_dispersion} and \eqref{eqn:extinction_P} only implies that the effective field~\eqref{eqn:effective_waves} solves~\eqref{eqn:ensemAsystem} for $X_1 > \bar X + \gamma R_o$. The missing information, needed to determine $\vec \alpha$, will come from solving~\eqref{eqn:ensemAsystem} for $0 \leq X_1 < \bar X + \gamma R_o$. We choose to do this by calculating a discrete solution for $\Ab_n$ within $0 \leq X_1 < \bar X + \gamma R_o$, and then matching the $\Ab_n$ with the effective waves~\eqref{eqn:effective_waves}.  The final result will be a (small) linear system~\eqref{eqn:matched-governing}.

\section{A one-dimensional integral equation}
\label{sec:finite_difference}
Due to the symmetry between the halfspace and the incident wave, we can reduce \eqref{eqn:ensemAsystem} to a one-dimensional Wiener-Hopf integral equation:
\begin{multline}
   \sum_{n=-\infty}^\infty
   \frac{\phi}{\pi R_o^2} \int_{0}^\infty T_{m} \A n (X_2) \psi_{n-m}(X_2-X_1) \mathrm d X_2
   \\
   \wrong{-}  \Ab_m(X_1)
    = - \ee^{\ii X_1 \cos \theta_\inc} T_{m} \ee^{\ii m ( \pi/2 - \theta_\inc )}, \quad \text{for} \;\; X_1 >0,
  \label{eqn:ensemAsystem_1D}
\end{multline}
where
\[
\psi_n(X) = S_{n}(X) + \chi_{\{ |X|< R_o \gamma\}}( B_{n}(X) - S_{n}(X)),
\]
with $\chi_{\{\mathrm{true}\}} = 1$ and $\chi_{\{\mathrm{false}\}} = 0$, $S_n(X)$ is given by~\eqref{eqn:Sn} and $B_n(X)$ is given by \eqref{eqn:B-approx}.

\edit{
Gerhard~\cite{kristensson_coherent_2015} deduced a similar one-dimensional integral equation for electromagnetism, and
in~\cite{gower_proof_2019} we showed that the analytic solution to~\eqref{eqn:ensemAsystem_1D} is a sum of effective plane waves.
}

We will use~\eqref{eqn:ensemAsystem_1D} to determine the effective waves~\eqref{eqn:effective_waves}, and to formulate a completely numerical solution to~\eqref{eqn:ensemAsystem}, which we use as a benchmark.

\subsection{The discrete form} \label{sec:discrete-form} The simplest discrete solution of~\eqref{eqn:ensemAsystem_1D} is to use a regular spaced finite difference method and a finite-section approximation\footnote{The kernel in~\eqref{eqn:ensemAsystem} does not satisfy the technical requirements in~\cite{de_hoog_finite-section_1987}, and we have been unable to find convergence results for approximating equations of the form~\eqref{eqn:ensemAsystem}. See~\cite{arens_solvability_2003} for a review on solvability.}. A similar finite difference solution was used in~\cite{gustavsson_multiple_2016}.

Let $\A n^j = \A n (X^j)$ for $X^j = j h$ and $j=0,\ldots,J$, with analogous notation for the other fields. We also define the vectors
\begin{align}
  & \vec \Ab_n = [\Ab_n^0,\Ab_n^1, \ldots, \Ab_n^J], \quad
  \vec b_n = -  \ee^{\ii n ( \pi/2 - \theta_\inc )} T_n [\ee^{ \ii X^0 \cos \theta_\inc}
  ,\ldots, \ee^{ \ii X^J \cos \theta_\inc}].
  \label{eqns:discretisation}
\end{align}
For implementation purposes, we consider all vectors to be column vectors. We also use the block matrix $\mathbb A$ with components $\mathbb A_{n1} = \vec \Ab_{n}$, that is
\begin{equation}
  \mathbb A =
   [\ldots  \,\,, \vec \Ab_{-n} \,, \vec \Ab_{1-n}\,, \ldots \,, \vec \Ab_{n} \,, \vec \Ab_{n +1}, \ldots ],
  \label{eqn:Ab_block}
\end{equation}
so $\mathbb A$ can be viewed as a one column matrix. The goal is to solve for $\mathbb A$.

To discretise the integrals in~\eqref{eqn:ensemAsystem_1D}, we use $\int f(X) dX \approx \sum_j f(X^j) \sigma_j$, which in the simplest form is $\sigma_j = h$ for every $j$. Discretising the integrals in \eqref{eqn:ensemAsystem_1D}, then substituting~\eqref{eqns:discretisation}, and $X_1^j = X_2^j = jh$ for $j =0,1,\ldots, J$,  leads to
\begin{equation}
  \sum_{n}({\mathcal E}_{nm}^\ell + {\mathcal R}_{nm}^\ell) \wrong{-} \Ab_m^\ell +
     \sum_{n} \sum_{j = 0}^J \edit{Q_{mn}^{\ell j}} \Ab_n^j  =  b_m^\ell, \quad \text{for} \;\; \ell = 0,1, \ldots, J,
  \label{eqn:ensemAsystemN}
\end{equation}
 where $q = \lfloor R_o\gamma/h \rfloor$,
\begin{align}
  & Q_{mn}^{\ell j} =   \frac{\phi T_m}{\pi R_o^2} \sigma_j S_{n-m}^{j-\ell} +   \frac{\phi T_m}{\pi R_o^2} \sigma_{\ell j} (B_{n-m}^{j-\ell} - S_{n-m}^{j -\ell} ) \chi_{\{|j-\ell| \leq q\}},
  \label{eqn:Q}
  \\
  & \mathcal E_{nm}^\ell  = \frac{\phi T_m}{\pi R_o^2} \int_{X_2 \geq X^J} \A n (X_2) S_{n-m}(X_2-X^\ell) dX_2,
  \label{eqn:E}
  \\
  & \mathcal R_{nm}^\ell  =  \chi_{\{\ell > J - q\}} \frac{\phi T_m}{\pi R_o^2} \times
  \label{eqn:R}
  \\ \notag
   &\hspace{1.5cm}
     \int_{X^J}^{X^\ell + R_o\gamma} \A n (X_2) (B_{n-m}(X_2 - X^\ell,k) - S_{n-m}(X_2 - X^\ell) )dX_2.
\end{align}
The $\sigma_{\ell j}$ depend on $\ell$ because the discrete domain of integration $|j-\ell| \leq q$ changes with $\ell$, though the simplest choice would still be $\sigma_{\ell j} = h$.


If we did not include $\mathcal E_{nm}^\ell$ and $\mathcal R_{nm}^\ell$, then the solution of \eqref{eqn:ensemAsystemN} would represent the average wave in the layer $0 \leq X \leq X^J$. One method to calculate the solution for the whole half-space $X \geq 0$ is to extend $X^J$ until $\A n (X^J)$ tends to zero.
However, it is more computationally efficient to calculate
$\mathcal E_{nm}^\ell$ and $\mathcal R_{nm}^\ell$ by approximating $\Ab_n(X)$ as a sum of plane waves, as shown below.

\section{Matching the discrete form and effective waves}
\label{sec:matching}
\begin{figure}[htbp]
   \centering
   \includegraphics[width=0.6\linewidth]{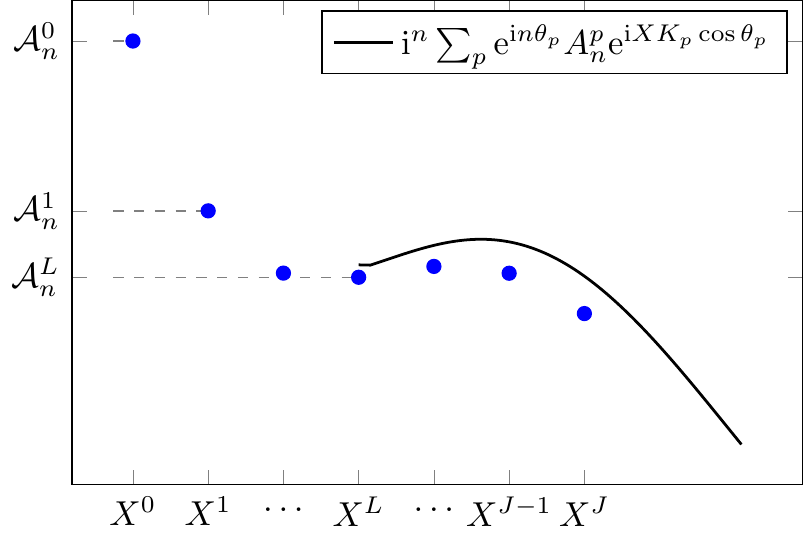}
   \caption{An illustration of the discrete solution $\Ab^j_n$~\eqref{eqns:discretisation} (blue circles) and the effective waves~\eqref{eqn:effective_waves} (black line).  We restrict the coefficients $A^p_n$ of the effective waves by imposing that the black line passes close to the $\Ab^j_n$ (i.e. satisfying the matching condiiton~\eqref{eqn:min_alpha}) for $X = X^L, \ldots, X^J$, where we chose $\bar X = X^L$. Increasing the number of effective waves will lead to a closer match between the discrete solution and effective waves.}
   \label{fig:graph-match}
 \end{figure}

Here we formulate a system to solve for the unknown effective wave amplitudes $\alpha^p$~\eqref{eqn:A_eigenvector} and $\mathbb A$~\eqref{eqn:Ab_block}. To do this, we substitute $\A n (X_2)$ for the effective waves~\eqref{eqn:effective_waves} in $ {\mathcal E}_{nm}^\ell$ and ${\mathcal R}_{nm}^\ell$~(\ref{eqn:E},\ref{eqn:R}),
 and we calculate the integral \edit{$g(\bar X)$} in~\eqref{eqn:extinction_P} by substituting $\A n (X_2)$ for the discrete solution $\Ab^j_n$~\eqref{eqns:discretisation}. Finally, to determine the $\alpha^p$, and therefore the effective waves~\eqref{eqn:effective_waves}, we impose that~\eqref{eqn:effective_waves} matches the discrete solution~\eqref{eqns:discretisation} in a thin layer near the boundary $\bar X$. For an illustration see \Cref{fig:graph-match}. Imposing this match acts like a boundary condition for the effective waves. From here onwards we assume that $\bar X = X^L$.

 \subsection{Using the effective waves to calculate~(\ref{eqn:E},\ref{eqn:R})}
  Substituting the effective waves~\eqref{eqn:effective_waves} into~\eqref{eqn:E}, then integrating and using~\eqref{eqn:Sn} we arrive at
 \begin{equation}
   \mathcal E_{nm}^\ell  =
      \\
      =  \frac{\phi T_m}{\pi R_o^2} \ii^{n+1} S_{n-m}^{J-\ell}
       \sum_{p=1}^P
       \frac{\ee^{  \ii X^J K_p  \cos \edit \varphi_p } \ee^{- \ii n \edit \varphi_p}}
       {K_p \cos \edit \varphi_p + \cos \theta_\inc} A_{n}^p,
     \label{eqn:E_eff}
 \end{equation}
 where we used $X^J - X^\ell = X^{J -\ell} \geq 0$, for $J \geq \ell$, when substituting $S_{n-m}(X^J-X^\ell)$ with~\eqref{eqn:Sn}.
 Employing~\eqref{eqn:A_eigenvector}, we write~\eqref{eqn:E_eff} in matrix form
 \begin{equation}
   \sum_n \mathcal E_{nm}^\ell = (\mathbf E_{m} \vec \alpha)_\ell, \quad
 (\mathbf E_{m})_{\ell p} =
   \frac{\phi T_m}{\pi R_o^2}  \sum_n \ii^{n+1} S_{n-m}^{J-\ell}
    \frac{\ee^{  \ii X^J K_p  \cos \edit \varphi_p } \ee^{- \ii n \edit \varphi_p}}
    {K_p \cos \edit \varphi_p + \cos \theta_\inc} a_{n}^p.
     \label{eqn:E_eff_matrix}
 \end{equation}

 To calculate~\eqref{eqn:R}, we first discretise the integral then substitute the effective waves~\eqref{eqn:effective_waves}, leading to
\begin{multline}
   \mathcal R_{nm}^\ell  =  \chi_{\{\ell > J - q\}} \frac{\phi T_m}{\pi R_o^2}
    \sum_{j=J}^{\ell + q} \A n (X^j) (B_{n-m}^{j-\ell} - S_{n-m}^{j-\ell} ) \sigma_{\ell j}
    \\
    = \chi_{\{\ell > J - q\}} \frac{\phi T_m \ii^n}{\pi R_o^2}
     \sum_{j=J}^{\ell + q}
      \sum_{p=1}^P  A_n^p \ee^{-\ii n \edit \varphi_p} \ee^{\ii X^j K_p \cos \edit \varphi_p}
      (B_{n-m}^{j-\ell} - S_{n-m}^{j-\ell} ) \sigma_{\ell j},
      \label{eqn:R_eff_components}
\end{multline}
 where $\sigma_{\ell j}$ represents the discrete integral in the domain $[X^J,X^{\ell + q}]$. Using~\eqref{eqn:A_eigenvector}, just as we did in~\eqref{eqn:E_eff_matrix}, we can write the above in a matrix form
 \begin{equation}
   \sum_n \mathcal R_{nm}^\ell  = (\mathbf R_{m} \vec \alpha)_\ell.
   \label{eqn:R_eff_matrix}
 \end{equation}

 We can now rewrite the integral equation~\eqref{eqn:ensemAsystemN}, using the above equations, in the compact form
 \begin{equation}
   (\mathbf E_{m} + \mathbf R_{m}) \vec \alpha \wrong{-} \vec I \vec \Ab_m
 + \sum_{n}  \vec Q_{mn} \vec \Ab_n
   =  \vec b_m,
   \label{eqn:matched-governing}
 \end{equation}
which is valid for all $m$. If $\vec \alpha$ was known, then we could calculate the discrete solution $\vec \Ab_n$ from the above. However, the $\vec \alpha$ also depends on the $\vec \Ab_n$, as we show below.

\subsection{The effective waves in terms of the discrete form} The equations to determine the effective waves, so far, are (\ref{eqn:Kp_dispersion}) and \eqref{eqn:extinction_P}. To calculate the integral in~\eqref{eqn:extinction_P}, we discretise and substitute~\eqref{eqns:discretisation},
which leads to the discrete form of the extinction equation~\eqref{eqn:extinction_P}:
\begin{equation}
 {\mathbf w^\mathrm{T}} \vec \alpha    = \mathbb G^\mathrm T \mathbb A  +  \ii \pi R_o^2 \cos \theta_\inc,
   \label{eqn:extinction_vector}
\end{equation}
where ${\cdot}^{\mathrm T}$ denotes the transpose, we used~\eqref{eqn:Ab_block}, $\edit{g(\bar X) =} \mathbb G^T \mathbb A = \sum_n \mathbf{G}_{n}^T \vec \Ab_n$,
\begin{align}
 & w^p  =   2 \phi \sum_{n = -\infty}^\infty  \ee^{\ii n  \theta_\inc}
   \ee^{-\ii n \edit \varphi_p}  \frac{\ee^{\ii (K_p \cos \edit \varphi_p- \cos \theta_\inc){X^L}}}{K_p \cos \edit \varphi_p- \cos \theta_\inc} a^p_n ,
 \\
 & (\mathbf{G}_{n})_{j} = 2 \phi \ee^{\ii n  \theta_\inc}   (-\ii)^{n-1} \ee^{-\ii X^j \cos \theta_\inc } \sigma_{j},
\label{eqn:extinction_vector_coefficients}
\end{align}
and as the domain of the integral in~\eqref{eqn:extinction_P} is only up to $X^L = \bar X \leq X^J$, we set $(\mathbf{G}_{n})_{j} = 0$ for $j > L$.

When using $P$ effective wavenumbers, there are $P$ unknowns $\alpha^1, \ldots, \alpha^P$, with, so far, only one scalar equation~\eqref{eqn:extinction_vector} to determine
them. To determine the $\alpha^p$, we match the sum of effective waves~\eqref{eqn:effective_waves} with the discrete form $\Ab_n^j$ in the interval: $X^L<X<X^J$, such as shown in~\Cref{fig:graph-match}. To do this we
 could enforce
\begin{equation}
\Ab_n^j = \ii^n \sum_p \ee^{-\ii n \edit \varphi_p} \ee^{\ii X^{j} K_p \cos \edit \varphi_p } a_{n}^p \alpha^p =  \vec \alpha^\mathrm{T} \mathbf v^j_n ,
\quad \text{for} \;\; j = L, L+1, \ldots, J.
  \label{eqn:match_mat}
\end{equation}
However, for $n \not =0$ the coefficients $\Ab_n^j$ and $a^p_n$ can be very small, and the above would not enforce the extinction equation~\eqref{eqn:extinction_vector}. So rather than use~\eqref{eqn:match_mat} for every $n$, it is more robust to
minimise the difference:
\begin{equation}
 \frac{1}{J-L} \min_{\vec \alpha} \sum_n \sum_{j=L}^{J} 
|\Ab_n^j - \vec \alpha^\mathrm{T} \mathbf v^j_n|^2 \quad \text{subject to} \quad
\mathbf w^\mathrm{T} \vec \alpha = \mathbb G^\mathrm{T} \mathbb A + \ii \pi R_o^2 \cos \theta_\inc,
\label{eqn:min_alpha}
\end{equation}
where the constraint enforces~\eqref{eqn:extinction_vector}.
For details on how to solve~\eqref{eqn:min_alpha} see \Cref{sec:complex_least_squares}. The solution to the above is
\begin{equation}
\vec \alpha = \mathbb L^\mathrm T \mathbb A + \frac{\ii \pi R_o^2 \cos \theta_\inc}{\mathbf w^T \mathbf V^{-1} \overline{\mathbf w}}\mathbf V^{-1} \overline{\mathbf w},
\label{eqn:alpha}
\end{equation}
where $\overline{\mathbf w}$ is the conjugate of $\mathbf w$, the block matrix $\mathbb L = [\ldots, \mathbf L_{-n}, \mathbf L_{1-n}, \ldots, \mathbf L_{n}, \ldots]$, with
\begin{align}
  & \mathbb L^\mathrm T \mathbb A = \sum_n \mathbf L^\mathrm T_n \vec \Ab_n, \quad
  \mathbf L^\mathrm T_n =  \mathbf Z^\mathrm T_n + w^{-1} \mathbf V^{-1} \overline{\mathbf w}  (\mathbf G_n^\mathrm T - \mathbf w^\mathrm{T} \mathbf Z^\mathrm T_n),
  \label{eqn:LT}
  \\
  & \mathbf V = \sum_n \sum_{j=L}^{J} \overline{\mathbf v}^j_n (\mathbf v^j_n)^\mathrm{T}, \quad
  \mathbf Z_{n}^\mathrm T = [\mathbf 0 \cdots \mathbf 0 \, \mathbf V^{-1}\overline{\mathbf v}_n^L \cdots \mathbf V^{-1}\overline{\mathbf v}_n^J].
\end{align}

Finally, substituting $\vec \alpha$~\eqref{eqn:alpha} into~\eqref{eqn:matched-governing} we reach an equation which we can solve for $\mathbb A$:
\begin{equation}
  \boxed{
    ( (\mathbb E + \mathbb R)\mathbb L^\mathrm T  + \mathbb M) \mathbb A  =  \mathbb B, \qquad \text{(Matching method)}
  }
\label{eqn:matched-governing-final}
\end{equation}
 where $\mathbb E$ and $\mathbb R$ have components $\mathbf E_{m}$ and $\mathbf R_{m}$, given by~\eqref{eqn:E_eff_matrix} and \eqref{eqn:R_eff_matrix}, respectively, while the components of block matrices $\mathbb B$ and $\mathbb M$ are
 \begin{align}
   & \mathbf B_m = \mathbf b_m - \frac{\ii \pi R_o^2 \cos \theta_\inc }{\mathbf w^T \mathbf V^{-1} \overline{\mathbf w}}(\mathbf E_m + \mathbf R_m) \mathbf V^{-1} \overline{\mathbf w},
   \\
 &\mathbf M_{mn} = \wrong{-}\delta_{mn}\mathbf I + \mathbf Q_{mn}.
   \label{eqn:ensemAsystemB}
 \end{align}

 To summarise, the terms $\mathbf w$, $\mathbf V$, and $\mathbb L$ are defined in the section immediately above, $\mathbf Q_{mn}$ is given by~\eqref{eqn:Q}, and both $\mathbb A$ and $\mathbf b_m$ are given by~\eqref{eqns:discretisation}. The angle $\theta_\inc$ is the angle of the incident plane wave~\eqref{eqn:incident}, $R_o$ is a non-dimensional particle radius~\eqref{eqns:non-dimensional} which increases with the frequency.
  The block matrices $\mathbb G$, $\mathbb B$, $\mathbb A$, $\mathbb E$, $\mathbb R$, $\mathbb L$, and $\mathbb Z$ all have only one column. The elements of these columns are either column vectors ($\mathbf G_{m}$, $\mathbf B_{m}$, $\vec \Ab_{m}$) or matrices ($\mathbf E_{m}$, $\mathbf R_{m}$, $\mathbf L_m$, and $\mathbf Z_m$).

\subsection{The matching algorithm} We can now understand how to truncate the effective wave series~\eqref{eqn:effective_waves}: assume the wavenumbers $K_p$ are ordered so that Im $K_p$ increases with $p = 1, \ldots, P$. Then note that the larger Im $(X^J K_p \cos \edit \varphi_p)$ the less the contribution this effective wave will make to the matching~\eqref{eqn:min_alpha}, $\mathbf w$~\eqref{eqn:extinction_vector}, $\mathcal R_{nm}^\ell$~\eqref{eqn:R_eff_components} and $\mathcal E_{nm}^\ell$~\eqref{eqn:E_eff_matrix}. That is, we can choose $P$ such that Im $(X^J K_P \cos \edit \varphi_P)$ is large enough so that this wave will not affect the solution $\mathbb A$.

\edit{To aid reproducibility, we explain how to solve equation~\eqref{eqn:matched-governing-final}, and determine $\mathbb A$, by using an algorithm in \Cref{sec:matching_algorithm}.}

\section{The resulting methods}
\label{sec:methods}
Here we summarise the Matching method, and other methods for solving~\eqref{eqn:ensemAsystem_1D}.
To differentiate between results for the different methods we use the superscripts $M$, $D$, and $O$. That is, we denote the field $\Ab_n(X)$ as
\begin{align}
  & \Ab_n^M(X) \quad (\text{Matching method}),  \quad  \Ab_n^D(X) \quad (\text{Discrete method}),
  \\
  & \Ab_n^O(X) \quad (\text{One-effective-wave method}).
\end{align}

For the Matching method, we solve~\eqref{eqn:matched-governing-final} to obtain
\begin{equation}
  \boxed{
    \Ab_{n}^M(X) =
    \begin{cases}
      \Ab_n^j = (\vec \Ab_n)_j &  X = X^j  \\
      \ii^n \sum_{p=1}^P \ee^{-\ii n \edit \varphi_p}\ee^{\ii X K_p \cos \edit \varphi_p} a_n^p \alpha^p &  X > X^J
    \end{cases} \; (\text{Matching method})
  }
  \label{eqn:matched-field}
\end{equation}
where the $\alpha^p$ are given from~\eqref{eqn:alpha}, $\edit \varphi_p$, $K_p$ and $a_n^p$ are solutions to (\ref{eqn:Snells}, \ref{eqn:Kp_dispersion}, \ref{eqn:A_eigenvector}). For details on the Matching method, see \Cref{alg:matched_method} in the supplementary material.

The one-effective-wave method is the typical method used in the literature. It consists in using only one effective wavenumber $K_1$, that is equation~\eqref{eqn:effective_wave} with $p=1$. This one wavenumber $K_1$ is often given explicitly in terms of either a low volume fraction or low frequency expansion. However, as we explore both moderate frequency and volume fractions, we will instead numerically solve for $K_1$, the least attenuating wavenumber. To solve for $K_1$ and $A_n^1$ we take $\bar X =0$ and numerically solve~\eqref{eqn:Kp_dispersion} and \eqref{eqn:extinction_P} for $P=p=1$. The Snell angle $\edit \varphi_1$ is determined from \eqref{eqn:Snells}, with $K=K_1$ and $\edit \varphi = \edit \varphi_1$. The result is
\begin{equation}
  \boxed{
    \Ab_{n}^O(X) = \ii^n \ee^{-\ii n \edit \varphi_1}\ee^{\ii X K_1 \cos \edit \varphi_1} A_n^1
    \quad (\text{One-effective-wave method})
  }
  \label{eqn:one-effective-field}
\end{equation}

From \Cref{sec:discrete-form}, we can devise a purely numerical method, which requires a much larger meshed domain for $X$. The resulting field is
\begin{equation}
  \boxed{
    \Ab_{n}^D(X^j) =
    \begin{cases}
      (\mathbb A^D_n)^j &  j \leq J\\
      0 &  j > J
    \end{cases} \quad (\text{Discrete method})
  }
\label{eqn:discrete-field}
\end{equation}
This discrete method gives a solution for a material occupying the layer $0 < X < X^J$ and $Y \in \mathbb R$. If the layer is deep enough, and the wave decays fast enough, then this discrete method will be the solution for an infinite half-space.
\Cref{alg:matched_method}, in the supplementary material, can be used to calculate this discrete method by taking $P=1$, $J =L$ instead of step 7, as there is no matching region, and replace steps 9-15 with: solve for $\mathbb A$ by using $\mathbb M \mathbb A^D =  \mathbb B$ instead of~\eqref{eqn:matched-governing-final}.

\subsection{Reflection coefficient} The reflection coefficient $\mathfrak R$ is the key information required for many measurement techniques. We can compare the different methods for calculating the average wave by comparing their resulting reflection coefficient, which is much simpler than comparing the resulting fields $\Ab_n (X)$.

Consider a particulate material occupying the region $x >0$ and choose a point  $(x,y)$ to measure the reflection, with $x <0$, then the ensemble average reflection coefficient $\mathfrak R$ is such that
\begin{equation}
  \ensem{u(x,y)} = u_\inc(x,y)
   + \mathfrak R \ee^{\ii k( - x \cos \theta_\inc + y \sin \theta_\inc)}.
   \label{eqn:reflected_wave}
\end{equation}
By combining (\ref{eqn:AverageWave} - \ref{eqn:u_to_A}), we conclude that
 \begin{equation}
   \mathfrak R =  \frac{\phi}{\pi R_o^2}  \ee^{\ii X \cos \theta_\inc }  \sum_n \int_{0}^\infty
   \Ab_n (X_1)  \int_{-\infty}^\infty \ee^{\ii Y_0 \sin \theta_\inc}
   F_{n}(\mathbf X_0)
   dY_0 dX_1,
 \end{equation}
where we used $\mathbf X_0 = \mathbf X_1 - \mathbf X$ and the non-dimensional parameters~\eqref{eqns:non-dimensional}. The integral in $Y_0$ is given by~\eqref{eqn:Sn} which, noting that $X_0 > 0$, leads to
\begin{equation}
  \mathfrak R =  \frac{2 \phi}{\pi R_o^2 \cos \theta_\inc}  \sum_n \ii^n \ee^{-\ii n \theta_\inc}\int_{0}^\infty
  \Ab_n (X_1)  \ee^{ \ii X_1  \cos \theta_\inc} dX_1.
  \label{eqn:reflection_coefficient}
\end{equation}
Substituting the Matching method field~\eqref{eqn:matched-field} into~\eqref{eqn:reflection_coefficient} leads to
\begin{equation}
    \boxed{
    \begin{aligned}
      &\mathfrak R^M =  \sum_{n=-\infty}^\infty \frac{2 \phi}{\pi R_o^2 \cos \theta_\inc}
      \times \hspace{4cm} (\text{Matching method})
      \\
      & \qquad \left[
      \ii^n \sum_{j=0}^J \sigma_j \A n^j \ee^{\ii X^j \cos \theta_\inc -\ii n \theta_\inc}
      +   \ii \sum_{p=1}^P \alpha^p a_{n}^p \ee^{\ii n \edit \varphi_\mathrm{ref}^p} \frac{\ee^{\ii X^J (K_p \cos \edit \varphi_p + \cos \theta_\inc)}}{K_p \cos \edit \varphi_p + \cos \theta_\inc}
      \right]
    \end{aligned}
    }
    \label{eqn:matched-reflection}
\end{equation}
where $\edit \varphi_\mathrm{ref}^p = \pi -\theta_\inc - \edit \varphi_p$. For an interpretation of the reflection angle $\edit \varphi_\mathrm{ref}^p$, see \cite[Figure 7]{gower_reflection_2018}.

For the discrete method, we discretise~\eqref{eqn:reflection_coefficient}, which leads to
\begin{equation}
  \boxed{
    \mathfrak R^D =  \sum_{n=-\infty}^\infty \frac{2 \phi}{\pi R_o^2 \cos \theta_\inc}
    \ii^n \sum_{j=0}^J \sigma_j \A n^j \ee^{\ii X^j \cos \theta_\inc -\ii n \theta_\inc}
    \quad (\text{Discrete method})
  }
  \label{eqn:discrete-reflection}
\end{equation}
Alternatively, to obtain the reflection coefficient for one effective wave~\eqref{eqn:one-effective-field}, we set $J=0$ and $P=1$ in \eqref{eqn:matched-reflection} to reach
\begin{equation}
  \boxed{
    \mathfrak R^O =  \sum_{n=-\infty}^\infty \frac{2 \phi}{\pi R_o^2 \cos \theta_\inc}
     \frac{\ii A_{n}^1 \ee^{\ii n \edit \varphi_\mathrm{ref}^1} }{K_1 \cos \edit \varphi_1 + \cos \theta_\inc}
    \;\; (\text{One-effective-wave method})
  }
  \label{eqn:one-effective-reflection}
\end{equation}
which agrees with equations (41) and (42) from \cite{martin_multiple_2011}, when expanding for low volume fraction $\phi$.

\section{Numerical experiments}
\label{sec:experiments}
For simplicity, we consider circular particles~\eqref{eqn:circular_t-matrix} for all numerical experiments, in which case, the non-dimensional radius~\eqref{eqns:non-dimensional} $R_o = a_o k$, where $a_o$ is the particle radius.

For the material properties we use a background material filled with particles which either strongly or weakly scatter the incident wave given, respectively, by
\begin{align}
& \frac{c_o}{c} = 0.5, \quad \frac{\rho_o}{\rho} = 0.5, \qquad \text{(strong scatterers)}
\label{eqn:strong-scatterers}
\\
& \frac{c_o}{c} = 1.1, \quad \frac{\rho_o}{\rho} = 8.0,  \qquad \text{(weak scatterers)}
\label{eqn:weak-scatterers}
\end{align}
noting that $\rho_o \gg \rho$ leads to weaker scattering than $\rho_o \ll \rho$.
We will use a range of angles of incidence $\theta_\inc$, particle volume fractions $\phi$, and particle radiuses $R_o$, which is equivalent to varying the incident wavenumbers $k$.

\subsection{Comparing the fields} 
\begin{figure}[h!]
  \centering
  \begin{tikzpicture}
  \node[inner sep=0pt] (p1) at (0,0)
      {\includegraphics[width=0.8\linewidth]{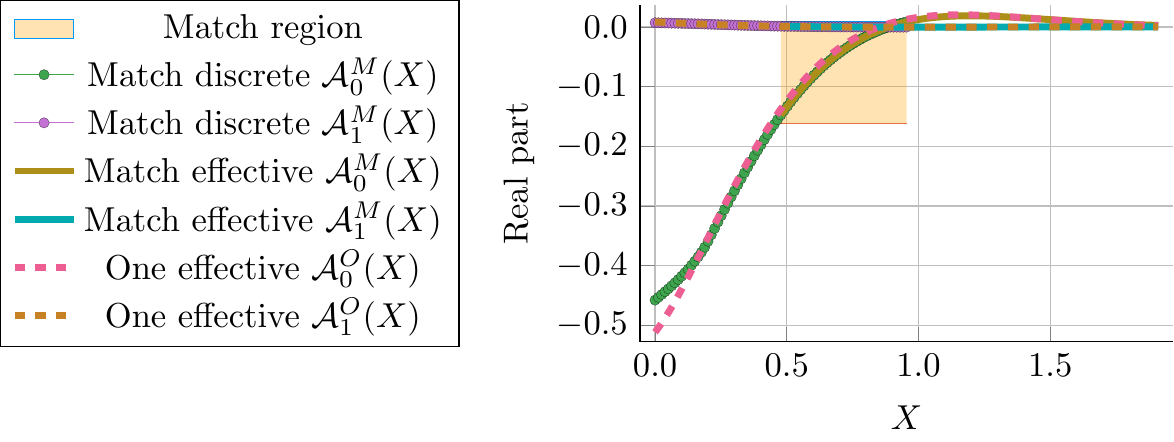}};
  \node[inner sep=0pt] (p2) at (-3.7,-4.2)
      {\includegraphics[width=0.45\linewidth]{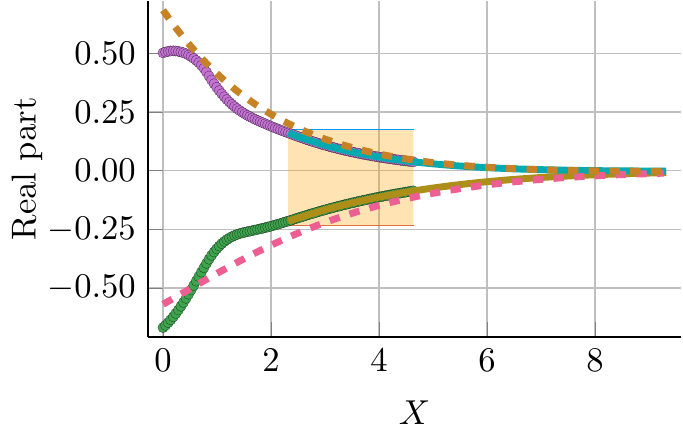}};
  \node[inner sep=0pt] (p3) at (2.4,-4.2)
      {\includegraphics[width=0.45\linewidth]{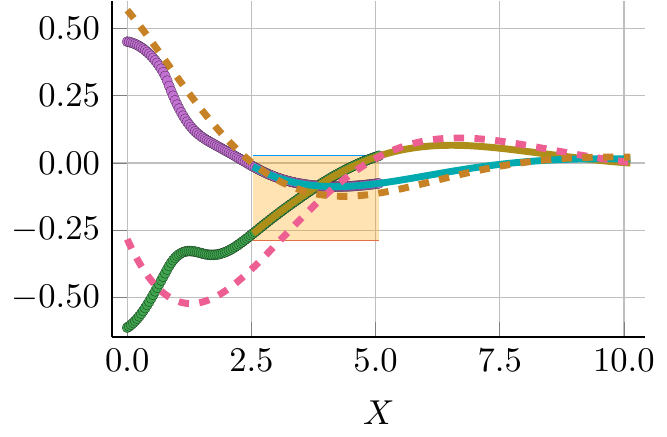}};
  \node[draw, fill=white] at (3.4,0.0) {$\phi = 8\%$, $R_o = 0.1$};
  \node[draw, fill=white] at (-2.3,-2.7) {$\phi = 14\%$, $R_o = 0.4$};
  \node[draw, fill=white] at (3.5,-2.7) {$\phi = 20\%$, $R_o = 0.4$};
  \end{tikzpicture}
  \caption{These graphs show the matching field~\eqref{eqn:matched-field} and the one-effective-wave field~\eqref{eqn:one-effective-field} for a material with circular particles, incident wave angle $\theta_\inc = 0$, and properties~\eqref{eqn:strong-scatterers}. The non-dimensional radius $R_o = k a_o$ and volume fraction $\phi$ are shown on each graph. We used six effective wavenumbers ($P =6$) for the bottom two graphs, and four effective wavenumbers ($P=4$) for the top right graph. Note that the discrete and effective part of the matching fields overlap in the match region.
   The one-effective-wave field in general loses accuracy close to the interface $X=0$, which is why it gives inaccurate predictions for the reflection coefficient $\mathcal R^O$~\eqref{eqn:one-effective-reflection}.}
  \label{fig:match_waves}
\end{figure}
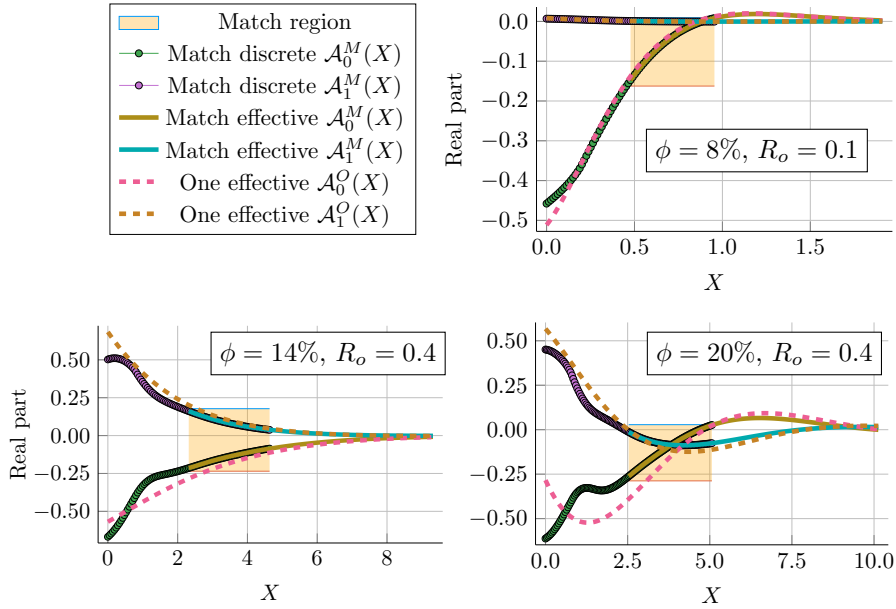

\Cref{fig:match_waves} shows several examples of $\Ab_n^M$ from~\eqref{eqn:matched-field}. As a comparison we have shown the one-effective-wave field $\Ab_n^O$~\eqref{eqn:one-effective-field} as well. To not clutter the figure, we have not shown the discrete field $\Ab_n^D$~\eqref{eqn:discrete-field}, which would lie exactly on top of $\Ab_n^M$.
\Cref{fig:match_waves} reveals how the discrete and effective wave parts of $\Ab_n^M$ very closely overlap in the matching region $X^L \leq X \leq X^J$. This close overlap is not due to over-fitting, as there are more than double the number of equations than unknowns.

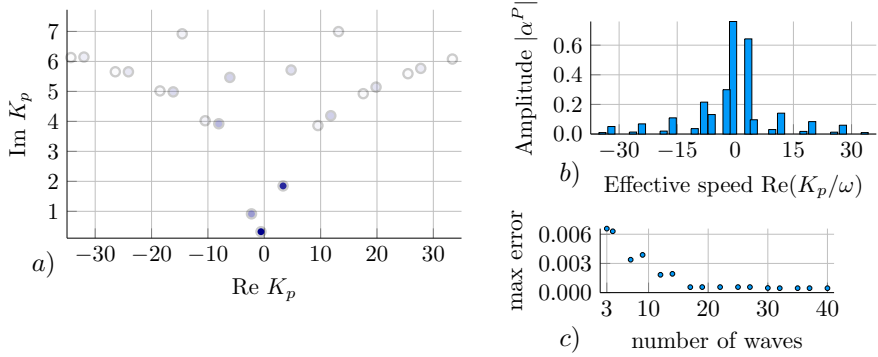
\begin{figure}[h!]
  \centering
  \begin{tikzpicture}
  \node[inner sep=0pt] (p2) at (-3.7,-4.4)
      {\includegraphics[width=0.47\linewidth]{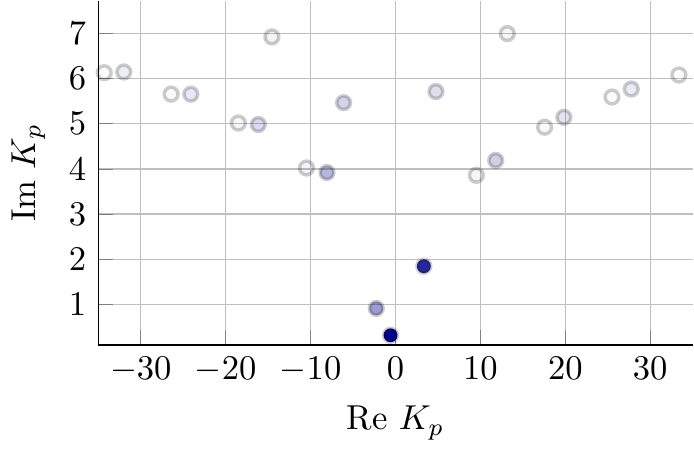}};
  \node[inner sep=0pt] (p3) at (2.4,-3.6)
      {\includegraphics[width=0.38\linewidth]{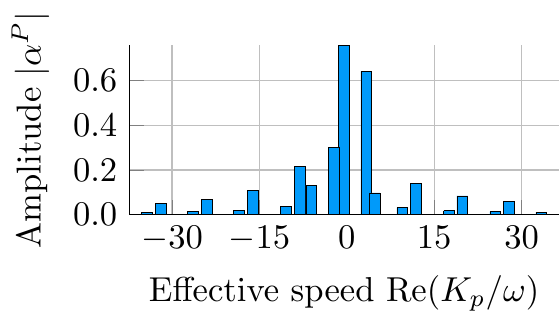}};
  \node at (-6.2,-5.8) {$a)$};
  \node at (0.8,-4.6) {$b)$};
  \node[inner sep=0pt] (p3) at (2.2,-6.0)
      {\includegraphics[width=0.35\linewidth]{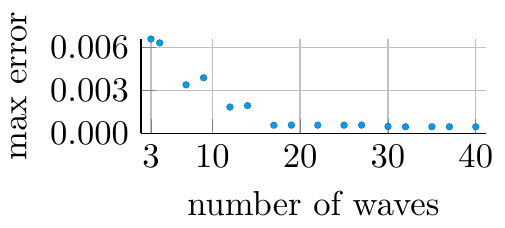}};
  \node at (0.8,-6.8) {$c)$};
  \end{tikzpicture}
  \caption{These graphs show the influence of the effective wavenumbers for the strong scatterers~\eqref{eqn:strong-scatterers} with particle volume fraction $\phi = 20\%$, non-dimensional radius $R_o = 0.4$, and incident wave angle $\theta_\inc = 0.4$. The resulting field $\Ab_n^M$ is shown in Figures~(\ref{fig:strong_match}).
  $a)$ shows the effective wavenumbers, with each marker corresponding to one wavenumber $K_P$ and its colour is stronger the larger the amplitude of its wave field $\alpha^P$. Clearly the larger the attenuation Im $K_p$, the lower the amplitude $\alpha^P$. $b)$ reveals how the amplitude $\alpha^P$ decreases when the effective phase speed increases in magnitude. $c)$ shows how the maximum error between the fields of the matching and discrete methods decrease when increasing the number of effective waves used by the Matching method. Note, if we had not included the three lowest attenuating wavenumbers, the maximum error would be larger than $0.17$.   }
  \label{fig:strong_wave_effs}
\end{figure}

\begin{figure}[h!]
  \centering
  \begin{tikzpicture}
    \node[inner sep=0pt] at (0.0,0.0)
        {\includegraphics[width=0.8\linewidth]{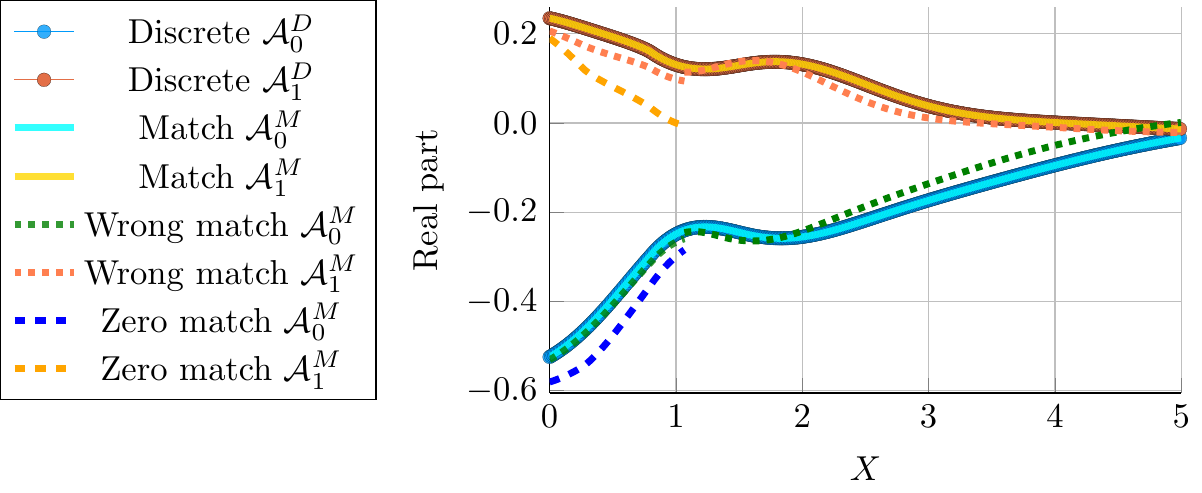}};
    \node[draw, fill=white] at (4.2,1.8) {$\phi = 20\%$, $R_o = 0.4$};
  \end{tikzpicture}
  \caption{This graph shows that the Matching method~\eqref{eqn:matched-field} overlaps with the discrete method~\eqref{eqn:discrete-field} (a purely numerical method). The effective wavenumbers used are shown in~\Cref{fig:strong_wave_effs}, and the material properties are given by~\eqref{eqn:strong-scatterers}. The dashed and dotted curves also result from the Matching method, but use the wrong effective waves: the dotted curve, wrong match $\Ab^M_0$ and $\Ab^M_1$, use the effective wavenumbers~\eqref{eqn:Kp_dispersion} multiplied by $1.2$. The zero-matching fields zeros all the effective wave amplitudes $a_n^p = 0$ and $\Ab^M_n(X) = 0$ for $X > 1$.}
  \label{fig:strong_match}
\end{figure}

We now look closely at a specific case: particle volume fraction $\phi = 20\%$ and non-dimensional particle radius $R_o = 0.4$ for the strong scatterers~\eqref{eqn:strong-scatterers}. \Cref{fig:strong_wave_effs} shows the effective wavenumbers used and how the greater the attenuation Im $K_p$, the lower the resulting amplitude $|\alpha^P|$ of the effective wave, and therefore the less it contributes to the total transmitted wave. We also see in \Cref{fig:strong_wave_effs}$c$ how increasing the number of effective waves (while fixing everything else), results in a smaller difference between the fields of the matching and discrete methods. This clearly confirms that the field $\Ab_n$ is composed of these multiple effective waves. \Cref{fig:strong_match} shows how the Matching method~\eqref{eqn:matched-field} and the discrete method~\eqref{eqn:discrete-field} closely overlap with
\[
\max_{X,n}\limits \|\Ab^M_n(X) - \Ab^D_n(X)\| = 4.5 \times 10^{-4},
\]
which is similar to the matching error $4.7 \times 10^{-5}$, given by the sum~\eqref{eqn:min_alpha}${}_1$. The dotted and dashed curves in \Cref{fig:strong_match} demonstrate how the Matching method is only accurate when using the effective wavenumbers that satisfy~\eqref{eqn:Kp_dispersion}. This agreement between the matching and discrete methods is not isolated to specific material properties and frequencies; we have yet to find a case where the two methods do not show excellent agreement\footnote{Naturally, when the truncation error of the discrete method is very large, we found that the result did not agree with the Matching method. Note that the truncation error of the discrete method is large when $\Ab_n(X)$ is weakly attenuating when increasing $X$.}. Further, when increasing the number of effective wavenumbers $P$, and lowering the tolerance $tol$ in~\cref{alg:matched_method}, the two methods converge to the same solution, as indicated by \Cref{fig:strong_wave_effs}$c$. In this paper we will not explore this convergence in detail, but we will show that the two methods produce the same reflection coefficient for a large parameter range.

\subsection{Comparing reflection coefficients} The reflection coefficient $\mathfrak R$ is a simple way to compare the different methods in~\Cref{sec:methods}.
Many scattering experiments aim to estimate $\mathfrak R$~\cite{west_comparison_1994,weser_particle_2013}. The accuracy of estimating $\mathfrak R$ is also directly related to the accuracy of calculating the transmitted waves.
\begin{figure}[ht]
  \centering
  \begin{tikzpicture}
  \node[inner sep=0pt] (p2) at (0.0,0.0)
      {\includegraphics[width=0.7\linewidth]{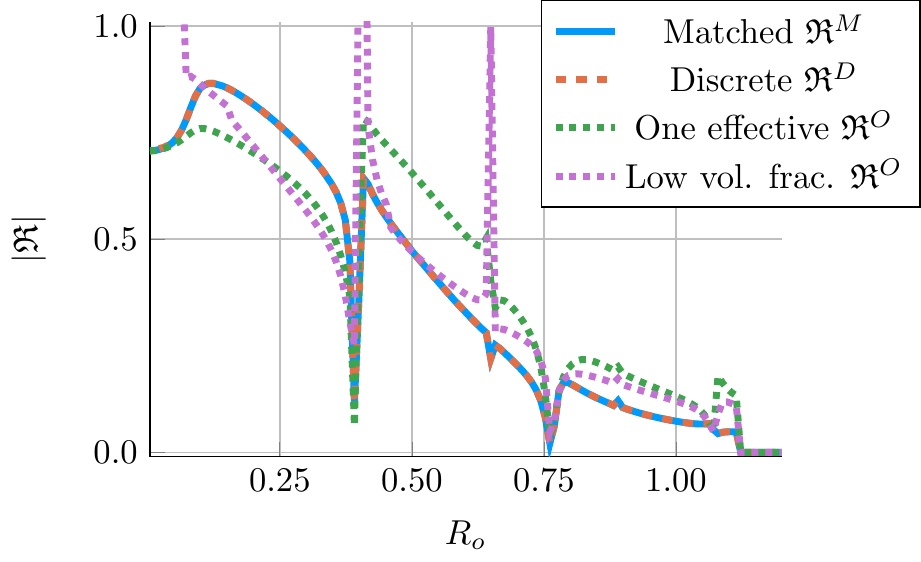}};
  \node[inner sep=0pt] (p3) at (-0.55,-5.5)
      {\includegraphics[width=0.614\linewidth]{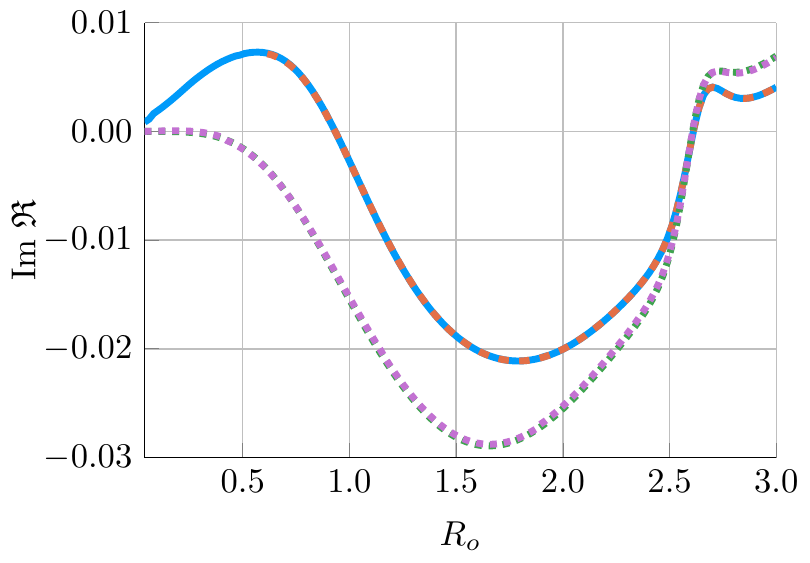}};
  \node at (-4.4,-2.0) {$a)$};
  \node at (-4.4,-7.8) {$b)$};
  \end{tikzpicture}
  \caption{ The reflection coefficients from the methods in~\Cref{sec:methods} as a function of the non-dimensional particle radius $R_o$: $a)$ has strong scattering particles~\eqref{eqn:strong-scatterers} with $\phi = 20 \%$ and $\theta_\inc = 0.0$ while $b)$ has weak scattering particles~\eqref{eqn:weak-scatterers} with $\phi = 25 \%$ and $\theta_\inc = 0.4$. Note the one-effective-wave fields almost overlap in this case. The real part of the curves in $b)$ are even closer together, with $\max_{R_o}\limits |\mathrm{Re}(\mathfrak R^O - \mathfrak R^M)| = 0.0026$ for the one effective $\mathfrak R^O$.  }
  \label{fig:reflection-coefficients}
\end{figure}

In \Cref{fig:reflection-coefficients} we compare the reflection coefficient for the discrete method $\mathfrak R^D$~\eqref{eqn:discrete-reflection}, Matching method
$\mathfrak R^M$~\eqref{eqn:matched-reflection} and two methods that use only one effective wavenumber~\eqref{eqn:one-effective-reflection}: \emph{one effective} $\mathfrak R^O$ uses a numerical solution for $K_1$ (the wavenumber with the smallest imaginary part), while the \emph{low vol. frac} $\mathfrak R^O$ uses a low-volume-fraction expansion for the wavenumber~\cite{martin_multiple_2011}.

In \Cref{fig:reflection-coefficients}$a$ we compare the reflection coefficients for strong scatterers~\eqref{eqn:strong-scatterers} when varying the particle radius $R_o$ (or likewise varying the wavenumber $k$) with a fixed volume fraction $\phi = 20\%$.
  We use at most 1600 points for the $X$ mesh, and less than 100 points for the $X$ mesh of the Matching method, and aim for a tolerance of $10^{-5}$ for the fields.

 We clearly see that $\mathfrak R^D$ and $\mathfrak R^M$~\eqref{eqn:matched-reflection} overlap. For $R_o > 0.03$ the maximum difference $\max_{R_o} |\mathfrak R^M - \mathfrak R^D| < 0.0014$. For $R_o < 0.03$ we have not shown $\mathfrak R^D$ because the numerical truncation error became too large (compared to our tolerance). This occurs when the fields $\Ab^D(X)$ decay slowly, which occurs for small particles (or low frequency). However, for low frequency the one effective $\mathfrak R^O$ is asymptotically accurate~\cite{parnell_multiple_2010}, and we see that $\mathfrak R^M$ does converge to $\mathfrak R^O$ as $R_o \to 0$. However,
 for larger $R_o$ the error of one effective $\mathfrak R^O$ is as much as $20\%$, while the low vol. frac. $\mathfrak R^O$ commits even larger errors. These larger errors are not unexpected, because the accuracy of the low-volume-fraction expansion depends on the type of scatterers and frequency~\cite{parnell_multiple_2010}, and can diverge in the limit $R_o \to 0$~\cite{gower_characterising_2018}.

\Cref{fig:reflection-coefficients}$b$ compares the reflection coefficients for weak scatterers~\eqref{eqn:weak-scatterers}.
We use at most 2200 points for the $X$ mesh, and less than 100 points for the $X$ mesh of the Matching method, and aim for a tolerance of $10^{-5}$ for the fields.

Again, as before, we do not show $\mathfrak R^D$ for values of $R_o$ where the numerical truncation error become large (relative to our tolerance). For this case of weak scatterers we see that the difference between the methods is less, though the reflection coefficient is also smaller with mean $|\mathfrak R^M| = 0.058$. Still, the relative error of Im $(\mathfrak R^M - \mathfrak R^O) \approx 10\%$. The imaginary part of the reflection coefficient, and where it changes sign, can be key for characterising random microstructure~\cite{roncen_bayesian_2018}. \edit{The real part of the reflection coefficients is not shown, as the relative errors for the real part are even smaller.}

\section{Conclusions}
\label{sec:conclusions}

Our overriding message is that there is not one, but a series of waves, with different effective wavenumbers, that propagate (with attenuation) in an ensemble averaged random particulate material. These waves must be included to accurately calculate reflection and transmission. \Cref{fig:wavenumbers} shows examples of these effective wavenumbers.

\edit{Although there is an analytic proof~\cite{gower_proof_2019} that there exists a series of effective waves, which solve  the equations~\eqref{eqn:ensemAsystem}, this current paper shows how to calculate these by using a Matching method~\eqref{eqn:matched-field}. In our numerical experiments in \Cref{sec:experiments}, we show that the Matching method converges to a numerical solution (the discrete method) for a broad range of wavenumbers $k$ (or equivalently the non-dimensional radius $R_o$), particle volume fractions, and two sets of material properties. For examples, \Cref{fig:strong_match} compares the average fields $\Ab_n(X)$,
and \Cref{fig:reflection-coefficients} the reflection coefficients $\mathfrak R$ of the matching and discrete methods.
The drawback of the discrete method~\eqref{eqn:discrete-field} is that it is computationally intensive, especially for low wave attenuation, requiring a spatial mesh between 1600 to 2000 elements to reach the same tolerance as the Matching method which used only 100 elements.}

\edit{For small incident wavenumbers $k$, the Matching method converges to a result which assumes there exists only one effective wave for both strong and weak scatterers.
Qualitatively, the fields $\Ab_n(X)$ from the one-effective-wave~\eqref{eqn:one-effective-field} and Matching method~\eqref{eqn:matched-field} agreed well when moving away from the material's interface, for example see~\Cref{fig:match_waves}. However, as the fields are not the same near the interface, the resulting reflection coefficients can significantly differ, as shown in \Cref{fig:reflection-coefficients}.}

\subsection{The next steps} \edit{Here we comment on a few directions for future work.}
 One important limit, that we did not investigate here, is the low-volume fraction limit: $\phi \ll 1$. \edit{In numerical experiments, not reported here, we found that the Matching method converges to the one-effective-wave method in the limit for low $\phi$.} It appears that as $\phi$ decreases the Im $K_p$, for $p>2$, tends to $+\infty$, implying that the boundary layer $\bar X$ shrinks and makes all but $K_1$ insignificant. This limit deserves a detailed analytic investigation in a separate paper.

The consequences of this work directly impacts upon effective wave methods used for acoustic, elastic, electromagnetic, and even quantum wave scattering. That said, many of these fields use vector wave equations and require the average intensity. So one challenge is to translate the results of this paper to vector wave equations and the average intensity. \edit{Note that for electromagnetic waves, much of the groundwork for the average fields has already been done~\cite{kristensson_coherent_2015,kristensson_evaluation_2015}.}

The radiative transfer equations are one outcome of properly deducing the averaged intensity for waves in particulate materials. For example, for electromagnetic waves, radiative transfer equations have been deduced under assumptions such as weak scattering, sparse particle volume fractions, and one effective wavenumber $K_1$~\cite{mishchenko_first-principles_2016}. Within the confines of the assumptions used, radiative transfer methods (and modifications) are leading to accurate predictions of the reflected intensity~ \cite{muinonen_coherent_2012,wegler_modeling_2006,przybilla_radiative_2006}. We speculate that this work will eventually lead to accurate predictions for reflected intensity for a broad range of frequencies and particles properties.

\section*{Data and reproducibility} All results can be reproduced with the publicly available software~\cite{gower_effective_waves.jl:_2018}, which has examples on how to calculate \href{https://github.com/arturgower/EffectiveWaves.jl/tree/v0.2.0/examples/many_wavenumbers}{the effective wavenumbers} and the \href{https://github.com/arturgower/EffectiveWaves.jl/tree/v0.2.0/examples/matched_method}{Matching met}, as well as the finite difference method we present.


\appendix

\section{Wiener-Hopf kernel}
Here we reduce \eqref{eqn:ensemAsystem} to the Wiener-Hopf equation~\eqref{eqn:ensemAsystem_1D}. First we separate the double integral:
\begin{multline*}
     \int_{\stackrel{x_2 > 0 }{\|\mathbf x_2-\mathbf x_1 \| > a_{12}}} \A n (k x_2) \ee^{\ii (y_2 -y_1) k \sin\theta_\inc} F_{n-m}(k\mathbf x_2 - k \mathbf x_1) \mathrm d \mathbf x_2 =
  \\
   \frac{1}{k^2} \int_{x_2>0} \A n (X_2) \int_{Y^2 > R_o^2 \gamma^2 - X^2}
   \ee^{\ii Y \sin\theta_\inc} F_{n-m}(\mathbf X) \mathrm d Y \mathrm d X,
\end{multline*}
where we used $\mathbf X = k\mathbf x_2 - k\mathbf x_1$ and the parameters~\eqref{eqns:non-dimensional}. We can then rewrite
\begin{multline}
  \int_{Y^2 > R_o^2 \gamma^2 - X^2} \ee^{\ii Y \sin\theta_\inc} F_{n-m}(\mathbf X) \mathrm d Y
 = \\
  \chi_{\{ |X|< R_o \gamma\}} B_{n-m}(X)
   +  \chi_{\{ |X|> R_o \gamma\}}S_{n-m}(X),
   \label{eqn:integral_F}
\end{multline}
where $\chi_{\{\mathrm{true}\}} = 1$ and $\chi_{\{\mathrm{false}\}} = 0$. From \cite[Eq. (37)]{martin_multiple_2011} we have
\begin{equation}
  S_n(X) =  \int_{-\infty}^\infty \ee^{\ii Y \sin\theta_\inc}
  F_{n}(\mathbf X) \mathrm d Y =
\frac{2}{\cos \theta_\inc}
\begin{cases}
  \ii^{n} \ee^{-\ii n \theta_\inc} \ee^{\ii X \cos \theta_\inc } & X \geq 0,\\
  (-\ii)^{n} \ee^{\ii n \theta_\inc} \ee^{-\ii X \cos \theta_\inc } & X < 0.
\end{cases}
  \label{eqn:Sn}
\end{equation}

The $B_{n-m}(X)$ in~\eqref{eqn:integral_F} only need to be evaluated for a small portion of the domain of $X$, and are given by
\begin{multline}
  B_n(X) = \int_{-\infty}^{\infty} \chi_{\{ Y^2> R_o^2 \gamma^2 - X^2\}} \ee^{\ii Y \sin\theta_\inc}
   F_{n}(\mathbf X) \mathrm d Y
  \\
  = 2(-1)^{n} \int_{\sqrt{R_o^2 \gamma^2 - X^2}}^\infty \cos \left( Y \sin\theta_\inc +  n \Theta\right) H_{n}(R) \mathrm d Y.
\end{multline}
Because the integrand tends to zero slowly as $Y$ increases, we use an asymptotic approximation to evaluate the integral, namely
\begin{align}
  & \cos \left( Y \sin\theta_\inc +  n \Theta\right) = \cos((n \pi)/2 + Y \sin(\theta_\inc)) + \mathcal O(X/Y),
  \\
  & H_{n}(R) =  -(-1)^{3/4} \ee^{-\ii n \pi /2} \sqrt{\frac{2}{\pi Y}} + \mathcal O(X^{3/2}/Y^{3/2}),
\end{align}
to rewrite
\begin{multline}
  B_n(X) = 2(-1)^{n} \int_{\sqrt{R_o^2 \gamma^2 - X^2}}^{Y_1} \cos \left( Y \sin\theta_\inc +  n \Theta\right) H_{n}(R) \mathrm d Y +
  \\ \frac{(1 + \ii)\ee^{\ii  Y_1 (1 - \sin \theta_\inc)}}{\sqrt{\pi Y_1} \cos^2 \theta_\inc}
  \left[ (-1)^n \ee^{2 \ii Y_1 \sin \theta_\inc} (1 - \sin \theta_\inc) +1+ \sin \theta_\inc \right] + \mathcal O(X/Y_1),
  \label{eqn:B-approx}
\end{multline}
then as $X$ is bounded by $|X|< R_o \gamma$, we can choose $Y_1$ such that $X/Y_1$ is below a prescribed tolerance.

Substituting~(\ref{eqn:integral_F},\ref{eqn:Sn},\ref{eqn:B-approx}) into~\eqref{eqn:ensemAsystem} leads to the Wiener-Hopf integral equation~\eqref{eqn:ensemAsystem_1D}.

\bibliographystyle{siamplain}
\bibliography{Library,SharedMultipleScattering}

\end{document}